\documentclass[aps,nofootinbib,prd,preprintnumbers,showpacs]{revtex4}
\usepackage{eurosym}
\usepackage{amsfonts}
\usepackage{amsmath}
\usepackage{hyperref}
\usepackage{amssymb}
\usepackage[english]{babel}
\usepackage{graphicx}
\usepackage{epsfig}
\usepackage{bm}
\usepackage{tikz}
\usepackage{feynmp-auto}
\usetikzlibrary{calc}
\usetikzlibrary{positioning,arrows}
\usetikzlibrary{decorations.pathmorphing}
\usetikzlibrary{decorations.markings}
\newcommand{\hs}{\hspace*{0.3cm}}

\newcommand{\be}{\begin{equation}}
	\newcommand{\ee}{\end{equation}}
\newcommand{\bea}{\begin{eqnarray}}
	\newcommand{\eea}{\end{eqnarray}}
\newcommand{\ben}{\begin{enumerate}}
	\newcommand{\een}{\end{enumerate}}
\newcommand{\bit}{\begin{itemize}}
	\newcommand{\eit}{\end{itemize}}
\newcommand{\bde}{\begin{widetext}}
	\newcommand{\ede}{\end{widetext}}
\newcommand{\nn}{\nonumber}
\newcommand{\crn}{\nonumber \\}

\newcommand{\al}{\alpha}
\newcommand{\la}{\lambda}

\newcommand{\ga}{\gamma}
\newcommand{\va}{\varphi}
\newcommand{\om}{\omega}
\newcommand{\pa}{\partial}

\newcommand{\fr}{\frac}

\newcommand{\bc}{\begin{center}}
	\newcommand{\ec}{\end{center}}
\newcommand{\Ga}{\Gamma}

\newcommand{\De}{\Delta}
\newcommand{\ep}{\epsilon}

\newcommand{\eq}{\eqref}

\newcommand{\mathsym}[1]{{}}

\newcommand{\Binh}[1]{{\color{blue}#1}}

\newcommand{\ITAR}{ Institute of Theoretical and Applied Research, Duy Tan University, Hanoi 10000, Vietnam}
\newcommand{\FNS}{Faculty of Natural Science, Duy Tan University, Da Nang 55000, Vietnam}
\newcommand{\stai}{ Subatomic Physics Research Group,
		Science and Technology Advanced Institute,\\
		Van Lang University, Ho Chi Minh City 70000, Vietnam}
\newcommand{\steh}{  Faculty of Applied Technology, School of  Technology,  Van Lang University, Ho Chi Minh City 70000, Vietnam}

\begin{document}
	\title{Higgs sector phenomenology in the 3-3-1 model with an  axionlike particle }
	\author{V. H. Binh$^{a,b}$}
	\email{vhbinh@iop.vast.vn}	
	\author{D. T. Binh$^{c,d}$}
	\email{dinhthanhbinh3@duytan.edu.vn}
	\author{A. E. C\'arcamo Hern\'andez$^{e,f,g}$}
	\email{antonio.carcamo@usm.cl}
	\author{D. T. Huong$^{b}$}
	\email{dthuong@iop.vast.vn}
	\author{D. V. Soa$^h$}
	\email{dvsoa@hnmu.edu.vn}
	\author{H. N. Long$^{k,l}$}
	\email{hoangngoclong@vlu.edu.vn (corresponding author)}
		\affiliation{
		$^a$ Graduate University of Science and Technology,
		Vietnam Academy of Science and Technology,
		18 Hoang Quoc Viet, Cau Giay, Hanoi 10000, Vietnam\\
		$^b$ Institute of Physics,   Vietnam Academy of Science and Technology, 10 Dao Tan, Ba
		Dinh, Hanoi 10000, Vietnam\\
		$^c$ \ITAR \\
		$^d$ \FNS\\	
		$^e$ Universidad T\'{e}cnica Federico Santa Mar\'{\i}a, Casilla 110-V,
		Valpara\'{\i}so, Chile\\
		$^f$ Centro Cient\'{\i}fico-Tecnol\'ogico de Valpara\'{\i}so, Casilla 110-V, Valpara\'{\i}so, Chile\\
		$^g$ Millennium Institute for Subatomic physics at high energy frontier - SAPHIR, Fernandez Concha 700, Santiago, Chile\\
$^h$
   Faculty of Applied Sciences, University of Economics and  Technology for Industries,
456 Minh Khai, Hai Ba Trung, Hanoi 10000, Vietnam\\	
		$^k$ \stai \\
$^l$ \steh 
	}
	\date{\today }

	\begin{abstract}
The scalar sector of the 3-3-1 model  with an axionlike particle	is studied in detail. In the model under consideration,  there are two kinds of scalar fields: the bilepton scalars carrying lepton number two and the ordinary ones without lepton number. We show that there is no mixing among these two kinds of scalar fields.
We analyze in detail the $CP$-odd scalar sector of the  model to find the physical fields of the axionlike particle and a pseudoscalar with mass in the range 100 GeV to 1 TeV. The results are different from others which have been published before.  The $CP$-even scalar sector of the  model is analyzed as well. The results of our analysis of the scalar sector allow us to accommodate scalar masses in the n$100$ GeV-$1$ TeV region. 
Furthermore we analyze the implications of the model in several flavor changing neutral decays of the top quark as well as in rare top quark decays. Besides that, the leptonic decays of the SM like Higgs boson as well as the meson oscillations are also analyzed. Our numerical analysis show that the model under consideration is consistent with the experimental constraints imposed by these processes.  
		
	\end{abstract}
	
	
	\maketitle

	\section{\label{intro}Introduction}
Nowadays, it is well known that the standard model (SM)  has to be extended. Among the extended models of the SM,	the versions based  on the $\mbox{SU(3)}_C\times \mbox{SU(3)}_L\times \mbox{U(1)}_X$ gauge group (called 3-3-1 models in short)  \cite{ppf1,ppf2,ppf3,ppf4,flt1,flt2,flt3,flt4,flt5,flt6} are of interest  with the following intriguing features such as the explanation on the number of  fermion generations, the electric charge quantization \cite{chargeq1,chargeq2}, source of $CP$ violation \cite{$CP$s1,$CP$s2} as well as the automatic fulfillment \cite{pal}
 of the Peccei-Quinn symmetry \cite{pq1,pq2}.
The Peccei-Quin symmetry for the economical 3-3-1 model \cite{eco,eco1,eco2,eco3,eco4,eco5} are
discussed in  Refs.  \cite{pal2,pal3}. The models contain self-interacting dark matter \cite{sidm1,sidm2}

The models are classified by a parameter $\beta$ appearing in the electric charge operator
	\be Q = T_3 + \beta \,  T_8 + X
	\,,
	\label{eq1}
	\ee
where $T_3$ and $T_8$ are $SU(3)_L$ generators, $X$ is the $U(1)_X$ charge.	The 3-3-1 model with arbitrary beta is presented in Ref. \cite{beta} (see also \cite{beta2}).  There are two main versions of the 3-3-1 models. The first one is the minimal model with $\beta = \pm \sqrt{3}$ which requires three $SU(3)_L$ scalar triplets and one $SU(3)_L$ scalar sextet \cite{ppf1,ppf2,ppf3}. Moreover, this version has a Landau pole around 5 TeV leading to a loss of pertubartivity around that scale. There exists effort to solve this puzzle  \cite{mimln}. In the recent work \cite{{mimlp}}, the Landau pole,  in the minimal version  by addition of octet leptoquarks,   can be around 100 TeV. The second one is the model with $\beta = \pm \fr 1{\sqrt{3}}$ which just requires three $SU(3)_L$ scalar triplets to provide masses for all fermions and bosons \cite{flt1,flt2,flt3,flt4,flt5}. This kind of model is more attractive due to its simpler scalar content and its lack of Landau pole at the TeV scale.

About two decades ago, the axion have been introduced in the 3-3-1 models \cite{a1,a2,a3}. The new nice property of the 3-3-1 model is found in a recent paper \cite{jpf}, where the cosmological inflation, axionlike particle (ALP) and seesaw mechanism are simultaneously addressed with a minimal scalar content. However, the above-mentioned paper contains some mistakes and  does not address phenomenological aspects related with flavor changing neutral process such as the $t \rightarrow hu$, $t \rightarrow hc$, $t \rightarrow u\gamma$ and $t \rightarrow c\gamma$ decays as well as the $K^{0}-\bar{K}^{0}$, $B_{d}^{0}-\bar{B}_{d}^{0}$ and $B_{s}^{0}-\bar{B}_{s}^{0}$ meson oscillations whose explanations, analysis and discussions are the purpose of this work.
	
	\section{Brief review of the model}
	\label{model}
	
	\subsection{Particle content and discrete symmetries}
	\label{dis}
	
To provide masses for fermions and to account for the existence of the ALP, the scalar sector of the model requires three $SU(3)_L$ scalar triplets $\eta, \rho, \chi$ as well as an electrically neutral $SU(3)_L$ scalar singlet $\phi$.
The scalar content of the model with their corresponding $\mbox{SU(3)}_C\times \mbox{SU(3)}_L\times \mbox{U(1)}_X$ assignments are given by:
	\bea \chi^T &=&\left( \chi^{0}_1,\chi ^{-}_2,\chi
	_{3}^{0}\right) \sim \left(1,3,-\fr 1 3\right), \,
	\eta^T =\left( \eta^{0}_1,\eta^{-}_2,\eta
	_{3}^{0}\right)  \sim\left(1,3,-\fr 1 3\right),
	\crn
	\rho^T & = & \left(
	\rho^{+}_1,\rho_2 ^{0},\rho_{3}^{+}\right) \sim \left(1,3,
	\fr 2 3\right), \,\hs
	\phi  \sim  (1,1,0)\,.
	\label{eq2}\eea
To provide  masses for the fermions and gauge bosons,  the above scalar fields have vacuum expectation values (VEVs) as follows
	\bea \langle \chi \rangle  &=& \fr 1{\sqrt{2}}\left( 0 , 0, v_{\chi}\right)^T, \,\hs
	\langle \eta \rangle = \fr 1{\sqrt{2}}\, \left( v_\eta , 0, 0\right)^T
	\crn
	\langle \rho \rangle & = & \fr 1{\sqrt{2}}\, \left( 0, v_\rho, 0
	\right)^T , \,\hs
	\langle \phi \rangle  = \fr 1{\sqrt{2}}\, v_\phi\,.
	\label{eq2t}\eea
where the VEV  $v_{\chi}$ triggers the spontaneous breaking of the $SU(3)_L\times U(1)_X$ gauge symmetry down to the SM electroweak gauge group. The remaining $SU(3)_L$ scalar triplets $\eta$ and $\rho$ break the SM electroweak gauge group.
	
On the other hand, the fermion spectrum of the model and their $\mbox{SU(3)}_C\times \mbox{SU(3)}_L\times \mbox{U(1)}_X$ assignments are:
	\bea
	&&\psi_{aL} =\left( \nu_{a},e_{a},  (\nu^{c}_R)^a\right) _{L}^{T}\sim
	\left( 1,3,-1/3\right),\hs e_{aR}\sim \left( 1,1,-1\right),\crn
	&& N_{aR} \sim (1,1,0),
	Q_{3L} =\left( u_3,d_3,U\right) _{L}^{T}\sim \left(
	3,3,1/3\right),\crn
	&& Q_{n  L} =\left( d_{n  },-u_{n	},D_{n  }\right)_{L}^{T}\sim \left( 3,3^{\ast },0\right), \crn
	&& u_{aR}, U_R \sim \left(3,1,2/3\right),\hs  d_{aR}, D_{n R}
	\sim (3,1,-1/3), \label{spectrumofparticles}\eea 
where $n  = 1,2$ and $a=\{n,3\}$ are family indices.
The $U$ and $D$ are exotic quarks with ordinary electric charges, whereas $N_{aR}$ are right-handed Majorana neutrinos.
	
The typical trouble of the 3-3-1 model with $\beta =  \pm \fr 1{\sqrt{3}}$ is that there are two triplets $\eta$ and $\chi$ with identical quantum numbers by	$\mbox{SU(3)}_L\times \mbox{U(1)}_X$ gauge group leading to the term $\mu^2_{\eta\chi}\eta^\dag \chi$, which complicates the structure of the square scalar mass matrices,  thus making  the analysis of the scalar sector very tedious. To avoid this kind of terms, one imposes the $Z_2$ discrete symmetry under which the $SU(3)_L$ scalar triplets $\eta$ and $\chi$ have opposite numbers, as done in Ref.\cite{a2}.  To provide Dirac and Majorana mass terms for $\nu_L$ and $N_R$ we have the above described particle content, shown in Table \ref{qnumber}. The particle assignments under the $\mbox{SU(3)}_C\times \mbox{SU(3)}_L\times \mbox{U(1)}_X \times Z_{11}\times Z_2$ group are	summarized in Table \ref{qnumber}. Here we have used a notation	$\om_k \equiv e^{i 2\pi \fr{k}{11}}, k=0,\pm 1\cdots \pm 5$.
	\small{
		\begin{table}[th]
			\resizebox{8cm}{!}{
				\begin{tabular}{|c|c|c|c|c|c|c|c|c|c|c|c|c|c|c|}
					\hline
					& $Q_{n L}$ & $Q_{3L}$ & $u_{a R}$ & $d_{a  R}$ & $U_{3R}$ &$ D_{n R} $ & $\psi_{aL}$ & $
					e_{aR}$ & $N_{aR} $ & $\eta$ & $\chi $ & $\rho$ & $
					\phi$  \\ \hline
					$SU(3)_C$ & $\mathbf{3}$ & $\mathbf{3}$ & $\mathbf{3}$ & $\mathbf{3}$ & $%
					\mathbf{3}$ & $\mathbf{3}$ & $\mathbf{1}$ & $\mathbf{1}$ & $\mathbf{1}$ & $%
					\mathbf{1}$ & $\mathbf{1}$ & $\mathbf{1}$ & $\mathbf{1}$   \\ \hline
					$SU(3)_L$ & $\overline{\mathbf{3}}$ & $\mathbf{3}$ & $\mathbf{1}$ & $%
					\mathbf{1}$ & $\mathbf{1}$ & $\mathbf{1}$ & $\mathbf{3}$ & $\mathbf{1}$ & $%
					\mathbf{1}$ & $\mathbf{3}$ & $\mathbf{3}$ & $\mathbf{3}$ & $\mathbf{1}$  \\ \hline
					$U(1)_X$ & $0$ & $\fr 1 3$ & $\fr 2 3$ & $-\fr 1 3$ & $\fr{2
					}{3}$ & $-\fr 1 3$ & $-\fr 1 3$ & $-1$ & $0$ & $-
					\fr 1 3$ & $-\fr 1 3$ & $\fr 2 3$ & $0$
					\\ \hline
					$Z_{11}$ & $\om^{-1}_4$ & $\om_0$ & $\om_5$ & $\om_2$
					& $\om_3$ & $\om_4$ & $\om_1$ & $\om_3$ & $%
					\om^{-1}_5$ & $\om^{-1}_5$ & $\om^{-1}_3$ & $\om^{-1}_2$ & $\om^{-1}_1$  \\
					\hline
					$Z_2$ & $1$ & $1$ & $-1$ & $-1$ & $1$ & $1$ & $1$ & $-1$ &
					$-1 $ & $-1$ & $1$
					& $-1$ & $1$ \\ \hline
			\end{tabular}}
			\caption{$SU(3)_C\times SU(3)_L\times U(1)_X\times Z_{11}\times Z_2$ charge assignments of the particle content of the model.
				Here $a=1,2,3$ and $\al =1,2$.}
			\label{qnumber}
	\end{table}}\newline
	
From  Table \ref{qnumber}, one recognizes   that under $Z_2$ symmetry, the following fields are \textit{odd}
	\be (\eta\,, \rho\,, u_R \,,  d_{n R}, e_{n R}, N_R)\,\,\,\,\rightarrow \,\,\,\,
	-\, (\eta\,, \rho\,,  u_R\,,  d_{n R}, e_{n R}, N_R)\,.
\label{oct1}	\ee

\subsection{Yukawa couplings}
	\label{yuka}
	
With the above specified particle content, the following Yukawa interactions invariant under the $SU(3)_C\times SU(3)_L\times U(1)_X\times Z_{11}\times Z_2$ symmetry, arise \cite{jpf}:
	\bea
	-\mathcal{L}^{Y}&=&y_{1}\bar{Q}_{3L}U_{R}\chi +\sum\limits_{n,m =1}^{2}\left(	y_{2}\right) _{n,m }\bar{Q}_{n L}D_{m R}\chi ^{\ast	}\crn
	&&+ \sum\limits_{a=1}^{3}\left( y_{3}\right) _{3a}\bar{Q}_{3L}u_{aR}\eta
	+\sum\limits_{n =1}^{2}\sum\limits_{a=1}^{3}\left( y_{4}\right) _{n a}\bar{Q}_{n L}d_{aR}\eta ^{\ast }\crn
	&&+\sum\limits_{a=1}^{3}\left( y_{5}\right) _{3a}\bar{Q}_{3L}d_{aR}\rho
	+\sum\limits_{n =1}^{2}\sum\limits_{a=1}^{3}\left( y_{6}\right) _{na}\bar{Q}_{n L}u_{aR}\rho ^{\ast }\crn
	&&+\sum\limits_{a=1}^{3}\sum\limits_{b=1}^{3}g_{ab}%
	\bar{\psi}_{aL}e_{bR}\rho
	+\sum\limits_{a=1}^{3}\sum\limits_{b=1}^{3}\left( y_{\nu }^{D}\right) _{ab}%
	\bar{\psi}_{aL}\eta N_{bR}\crn
	&&+\sum\limits_{a=1}^{3}\sum\limits_{b=1}^{3}\left(
	y_{N}\right) _{ab}\phi \bar{N}_{aR}^{C}N_{bR}+\mbox{H.c.}.
	\label{yukintera}
	\eea
Let us note that the above given Yukawa interactions in (\ref{yukintera}) are invariant only under	the  $Z_2$ assignment given above.	
 It is emphasized that the transformation under the $Z_2$ in this paper is different from than the one given in Ref. \cite{jpf} where $\chi$ is odd.

The exotic quarks get masses from $v_{\chi}$, top quark get mass from $v_\eta$, charged leptons get masses from $v_\rho$, while new Majorana neutrino $N_R$ gets mass through $v_\phi$. The Dirac neutrino mass term arises from $v_\eta $, while the Majorana mass term arises from $v_\phi$ (see last two terms in \eq{yukintera}). From the last two terms of Eq. (\ref{yukintera}), it follows that the tiny masses for the light active neutrinos are generated from 	a type I seesaw mechanism mediated by right handed Majorana neutrinos, thus implying that the resulting light active neutrino mass matrix has the form:
	\be
	M_{\nu}=M^D_{\nu}M^{-1}_N\left(M^D_{\nu}\right)^T,\hs  M^D_{\nu}=y^{D}_{\nu}\fr{v_{\eta}}{\sqrt{2}},
	\hs  M_N=\sqrt{2}\,y_{N}v_{\phi}.
	\ee
			
	\subsection{Gauge bosons}
	\label{gboson}
	
First of all, the model has nine  electroweak gauge bosons arising from the $SU(3)_L\times U(1)_X$ symmetry. Their interactions with the $SU(3)_L$ scalar triplets are included in the following kinetic terms:
	\be \mathcal{L}_{Higgs} = \sum_{H=\chi, \eta,\rho,\phi} (D^\mu H)^\dag D_\mu H \,,
	\label{j231}
	\ee
where the covariant derivative is given by
	\be  D_{\mu}\equiv \pa _{\mu}-i g T^a W^a_{\mu}-i g_X X T^9X_{\mu},  \label{j232}\ee
where $T^9=1/\sqrt{6}I_{3\times 3}$ being $I_{3\times 3}$ the $3\times 3$ identity matrix and $g$, $g_X$ are gauge couplings of the two groups $SU(3)_L$ and $U(1)_X$, respectively. Secondly, the matrix $W^aT^a$, where $T^a =\la_a/2$ corresponds to a triplet representation, is written as follows:
	\bea W^a_{\mu}T^a=\frac{1}{2}\left(
	\begin{array}{ccc}
		W^3_{\mu}+\fr{1}{\sqrt{3}} W^8_{\mu}& \sqrt{2}W^+_{\mu} &  \sqrt{2}Y^{+}_{\mu} \\
		\sqrt{2}W^-_{\mu} &  -W^3_{\mu}+\fr{1}{\sqrt{3}} W^8_{\mu} & \sqrt{2}X^{0}_{\mu} \\
		\sqrt{2}Y^{-}_{\mu}& \sqrt{2}X^{0*}_{\mu} &-\fr{2}{\sqrt{3}} W^8_{\mu}\\
	\end{array}
	\right),
	\label{j233}\eea
in which we have defined the mass eigenstates of the charged gauge bosons as
	\bea W^{\pm}_{\mu } &= &\fr{1}{\sqrt{2}}\left( W^1_{\mu}\mp i W^2_{\mu}\right),\crn
	Y^{\pm }_{\mu} &= &\fr{1}{\sqrt{2}}\left( W^4_{\mu}\mp i W^5_{\mu}\right),\hs
	X^{ 0}_{\mu}=\fr{1}{\sqrt{2}}\left( W^6_{\mu}- i W^7_{\mu}\right),\hs
 X^{0*}_{\mu}=\fr{1}{\sqrt{2}}\left( W^6_{\mu}+i W^7_{\mu}\right) .
	\label{j234}\eea
	
After spontaneous  symmetry breaking, the mass spectrum of the gauge bosons arise from the following terms: \be \mathcal{L}_{mass} = \sum_{H=\chi,\eta,\rho}(D^\mu\langle H \rangle)^\dag (D_\mu \langle H\rangle)\,.\ee
 
The  charged and  {\it bilepton}  gauge bosons get masses given by:
	\be m^2_{W} = \fr{g^2}{4}(v_\eta^2+v_\rho^2),\hs m^2_{X^0}=\fr{g^2}{4}(v_{\chi}^2+v_\eta^2),\hs m^2_Y=\fr{g^2}{4}(v_{\chi}^2+v_\rho^2)\,.\label{j235} \ee
$W$ is identical to that of the standard model, while $(X,Y)$ form a new, heavy gauge vector doublet with a mass splitting \cite{il}
\[|m^2_Y-m^2_{X^0}|< m^2_W\,.\]

From \eq{j235}, it follows
	\be v_\eta^2+v_\rho^2 = v_{ew}^2 = 246^2 \,  \textrm{GeV}^2\,.
	\label{j236} \ee
	
Finally, there is a mixing among the $W_3, W_8, B$ components. In the basis of these elements, the mass matrix is given by
	\be M_{neural}^2=\fr{g^2}{4}\left(
	\begin{array}{ccc}
		v_\eta^2+v_\rho^2 & \fr{v_\eta^2-v_\rho^2}{\sqrt{3}} & -\fr{2t}{3\sqrt{6}}(v_\eta^2+2v_\rho^2)  \\
		\fr{v_\eta^2-v_\rho^2}{\sqrt{3}} & \fr{1}{3}(4v_{\chi}^2+v_\eta^2+v_\rho^2) &
		\fr{\sqrt{2}t}{9}(2v_{\chi}^2-v_\eta^2+2v_\rho^2)
		\\
		-\fr{2t}{3\sqrt{6}}(v_\eta^2+2v_\rho^2) & \fr{\sqrt{2}t}{9}(2v_{\chi}^2-v_\eta^2+2v_\rho^2)
		& \fr{2t^2}{27}(v_{\chi}^2+v_\eta^2+4v_\rho^2)
		\\
	\end{array}%
	\right), \label{nmass}\ee
where \be t=\fr{3\sqrt{2}s_W}{\sqrt{3-4s^2_W}}.\ee
	
Diagonalization proceeds through two steps, in the first step	the $3\times 3 $ matrix reduces to one block diagonalized which yields a $2\times 2$ matrix in the bottom.	The eigenstates are now rewritten as follows
\bea A_\mu &=& s_W
	W_{3\mu}+c_W\left(-\fr{t_W}{\sqrt{3}}
	W_{8\mu}+\sqrt{1-\fr{t^2_W}{3}}B_\mu\right),\crn
	Z_\mu&=& c_W
	W_{3\mu}-s_W\left(-\fr{t_W}{\sqrt{3}}
	W_{8\mu}+\sqrt{1-\fr{t^2_W}{3}}B_\mu\right),\crn Z'_\mu &=&
	\sqrt{1-\fr{t^2_W}{3}} W_{8\mu}+\fr{t_W}{\sqrt{3}}B_\mu.\eea
	
From the analysis of the gauge sector, we found one massless gauge boson, which corresponds to the  photon $A$. Furthermore, besides the bilepton gauge bosons, the neutral gauge boson spectrum contains two massive neutral gauge bosons $Z$ and $Z^\prime$. The elements of the neutral squared gauge boson mass matrix in the $(Z,Z^\prime)$ basis is given by
	\bea
	m^2_Z&=&\fr {g^2}{4c^2_W} (v_\rho^2+v_\eta^2),\label{j11}\\
	m^2_{ZZ'} &=& \fr{g^2\left[( t^2_W-1)v_\rho^2+( t^2_W+1)v_\eta^2\right]}{4\sqrt3 c_W\sqrt{1-\fr 1 3 t^2_W}},\\
	m^2_{Z'}&= &\fr{g^2\left[4v_{\chi}^2+( t^2_W-1)^2v_\rho^2+( t^2_W+1)^2v_\eta^2\right]}{4(3- t^2_W)}\, .\eea
	
Finally, this matrix is diagonalized by the following field transformations
	\bea
	Z^1_\mu  &= & c_{\theta_Z} {\mathcal Z}_{\mu}-s_{\theta_Z} \mathcal{Z}'_{\mu},\crn
	Z^2_\mu& = & s_{\theta_Z} {\mathcal Z}_{\mu}+c_{\theta_Z} \mathcal{Z}'_{\mu}\,
	\label{j251}\eea
where \cite{Aoidong}
	\bea
	t_{2 \theta_Z} &=&\fr{s_{\theta_Z}}{c_{\theta_Z}}=\fr{2m^2_{ZZ'}}{m^2_{Z'}-m^2_Z}\simeq \fr{\sqrt{(3- t^2_W)}\left[( t^2_W-1)v_\rho^2+( t^2_W+1)v_\eta^2\right]}{2v_{\chi}^2 c_W},\\
	m^2_ {Z_1} &=& \fr1 2 \left[m^2_Z +m^2_{Z'} - \sqrt{(m^2_Z - m^2_{Z'})^2 + 4m^4_ {ZZ'} }\right]
	\simeq m^2_Z-\fr{m^4_{ZZ'}}{m^2_{Z'}}\crn
	&\simeq& \fr{g^2}{4c^2_W}\left\{v_\rho^2+v_\eta^2-\fr{\left[( t^2_W-1)v_\rho^2+( t^2_W+1)v_\eta^2\right]^2}{4v_{\chi}^2}\right\} \approx \fr{ m_{W}^2}{c^{2}_{W}},\crn
	m^2_ {Z_2} &=& \fr1 2 \left[m^2_Z +m^2_{Z'} + \sqrt{(m^2_Z - m^2_{Z'})^2 + 4m^4_ {ZZ'} }\right]\simeq m^2_{Z'}  \approx  \fr{g^{2}c^{2}_{W}}{(3-4s_W^2)}v_{\chi}^2 .\label{j12}
	\eea
	
Note that exotic quarks $U$ and $D_\al$ as well as gauge bosons $X^0, Y^\pm$ carry lepton number two \cite{joshi,cl,jh18}.
The  gauge boson couplings of this model are the same  in Refs. \cite{self1,self2}.
Due to quark family  discrimination, there are flavor changing neutral currents mediated by $Z^\prime$ at the tree level \cite{fc1,fc2,fc3,lv}.

	\section{Higgs potential}
	\label{potential}
	
	The model scalar potential has the form:
	\bea
	V &=& \mu^2_\phi  \phi^* \phi +  \mu_\chi^2  \chi^\dag \chi + \mu_\rho^2
	\rho^\dag \rho +  \mu_\eta^2 \eta^\dag \eta + \la_1 ( \chi^\dag \chi)^2 + \la_2 ( \eta^\dag
	\eta)^2\crn &  & + \la_3 ( \rho^\dag \rho)^2 +
	\la_4 ( \chi^\dag \chi)( \eta^\dag \eta) + \la_5 ( \chi^\dag \chi)( \rho^\dag \rho) + \la_6 ( \eta^\dag \eta)( \rho^\dag \rho)\crn
	&& + \la_7 ( \chi^\dag \eta)( \eta^\dag \chi) + \la_8 ( \chi^\dag \rho)( \rho^\dag \chi) + \la_9 ( \eta^\dag \rho)( \rho^\dag \eta)\crn
	&& + \la_{10} ( \phi^* \phi)^2 + \la_{11} ( \phi^* \phi)( \chi^\dag \chi) + \la_{12} ( \phi^* \phi)( \rho^\dag \rho)\crn
	&& + \la_{13} ( \phi^* \phi)( \eta^\dag \eta)
	+ \left( \la_\phi\ep^{ijk} \eta_i \rho_j \chi_k \phi
	+H.c.\right) \label{poten3} \eea

The VEV $v_\phi$ is responsible for the PQ symmetry breaking resulting in the existence of invisible ALP due to very high scale around $10^{10}-10^{11}$ GeV. Then $\mbox{SU(3)}_L\times \mbox{U(1)}_X$ breaks to the SM group by $v_\chi$ and two others $v_\rho, v_\eta$ are needed for the usual $ \mbox{U(1)}_Q$	symmetry. Hence $ v_\phi \gg v_\chi \gg v_\rho, v_\eta$.	The constraint conditions of such scalar potential were analyzed in Ref. \cite{a2}. From \eq{poten3}, it is reasonable to assume: $\la_2 \approx \la_3, \, \la_4 \approx \la_5, \, \la_7 \approx \la_8, \,\la_{12} \approx \la_{13}$. According  Ref. \cite{padax}, $v_\chi \geq 10357$ GeV for $M_{Z'} \geq 4.1$ TeV.
	
Let us expand these scalar fields around their VEVs.
	\bea
	\rho_2^0 &= & \fr{1}{\sqrt{2}}(v_\rho + R_\rho + i I_\rho)\, , \hs \eta_1^0 = \fr{1}{\sqrt{2}}(v_\eta + R^1_\eta + i I^1_\eta)\, ,\crn
	\chi_3^0 &= & \fr{1}{\sqrt{2}}(v_\chi + R^3_\chi + i I^3_\chi)\, , \hs \phi  = \fr{1}{\sqrt{2}}(v_\phi + R_\phi + i I_\phi)\, . \label{poten4}
	\eea

Substitution of (\ref{poten4})  into  (\ref{poten3})  leads to the following constraints at the tree level as follows
	\bea
	\mu^2_\rho + \la_3 v_\rho^2 + \fr{\la_5}2 v_\chi^2 + \fr{\la_6}2 v_\eta^2 + \fr{\la_{12}}2 v_\phi^2 + \fr{A}{2 v_\rho^2} & = & 0\, ,\crn
	\mu^2_\eta + \la_2 v_\eta^2 + \fr{\la_4}2 v_\chi^2 + \fr{\la_6}2 v_\rho^2 + \fr{\la_{13}}2 v_\phi^2 + \fr{A}{2 v_\eta^2} & = & 0\, ,\crn
	\mu^2_\chi + \la_1 v_\chi^2 + \fr{\la_4}2 v_\eta^2 + \fr{\la_5}2 v_\rho^2 + \fr{\la_{11}}2 v_\phi^2 + \fr{A}{2 v_\chi^2} & = & 0\, ,\crn
	\mu^2_\phi + \la_{10}  v_\phi^2 + \fr{\la_{11}}2 v_\chi^2 + \fr{\la_{12}}2 v_\rho^2 + \fr{\la_{13}}2 v_\eta^2 + \fr{A}{2 v_\phi^2} & = & 0\, , \label{poten5}
	\eea
where $A \equiv \la_\phi v_\phi v_\chi v_\eta  v_\rho $.

	\subsection{Charged scalar sector}
	\label{potentialcc}
There are four charged scalar fields: $\eta_2^-, \rho_1^-, \rho_3^-$ and $\chi_2^-$.

i)  In the basis ($\eta_2^-, \rho_1^- $), 
the corresponding squared mass matrix is given by:
	\be
	M_c = \left(
	\begin{array}{cc}
		\fr{ \la_9 v_\rho^2}{2} - \fr{A}{2 v^2_\eta} &  \fr{ \la_9 v_\rho v_\eta}{2} - \fr{A}{2 v_\rho v_\eta }\\
		\fr{ \la_9 v_\rho v_\eta}{2} - \fr{A}{2 v_\rho v_\eta} &  \fr{ \la_9 v_\eta^2}{2} - \fr{A}{2 v^2_\rho} \\
	\end{array}
	\right) = -\fr{ (A - \la_9 v_\rho^2 v^2_\eta)}{2} \left(
	\begin{array}{cc}
		\fr{1}{v_\eta^2} &  \fr{1}{v_\eta v_\rho}\\
		\fr{1}{v_\eta v_\rho} &  \fr{1}{v_\rho^2} \\
	\end{array}
	\right)\, .
	\label{c1}
	\ee
From this matrix, we get the  massless $G_1^{\pm}$ states and two massive ones, i.e., $H^{\pm}_1$ with mass equal to
	\be m^2_{H^{\pm}_1} =-\fr{ (A - \la_9 v_\rho^2 v^2_\eta)}{2} . \fr{(v^2_\rho +v^2_\eta )}{v^2_\rho v^2_\eta}
	\label{c2}
	\ee
Let us note that the $G_1^{\pm}$ massless charged scalar fields correspond to the SM charged Goldstone bosons associated with the longitudinal components of the $W^{\pm}$ gauge bosons.
	
The physical fields are given by
	\be
	\left(
	\begin{array}{c}
		G_1^{\pm} \\
		H^{\pm}_{1} \\
	\end{array}
	\right) = \left(
	\begin{array}{cc}
		\cos \al  & - \sin \al  \\
		\sin \al  &  \cos \al  \\
	\end{array}
	\right) \left(  \begin{array}{c}
		\rho_1^{\pm}\\
		\eta^{\pm} \\
	\end{array}
	\right)\, ,
	\label{c3}
	\ee
	where
	\be
	\tan  \al  =\fr{v_\eta}{ v_\rho}\, .
	\label{c4}
	\ee	
	From (\ref{c2}) it follows
	\be
	\la_9 > \la_\phi \fr{v_\phi v_\chi}{v_\rho v_\eta}  = \fr{A}{v^2_\rho v^2_\eta} =
 \fr{A}{( v_{ew}^2 - v^2_\eta) v^2_\eta}	\,.
	\label{kl1}
\ee
From \eq{kl1}, ones get  condition for  the perturbative coupling as follows   \be
\fr{\vert A \vert }{( v_{ew}^2 - v^2_\eta) v^2_\eta} < 1.
\label{t1}
\ee
	
Then, the constraint for the important coupling $\la_\phi$ is given by
\be 
	\vert A \vert < ( v_{ew}^2 - v^2_\eta) v^2_\eta \hs  \Rightarrow \hs \vert \la_\phi \vert < \fr{( v_{ew}^2 - v^2_\eta) \tan \alpha}{v_\phi v_\chi}\,.
\label{t4}
\ee	
For simplicity, let us assume $v_\eta = v_\rho =  v_{ew}/\sqrt{2} \simeq  174 $ GeV, $v_\phi = 10^{10}$ GeV and $v_\chi = 10^5$ GeV, then $\vert \la_\phi \vert < 10^{-11}$.  It is interesting to note that such tiny couplings (Yukawa couplings responsible for proton instability) arise also  in the supersymmetric 3-3-1 model \cite{pl}.

 ii) For the charged scalars, in the basis ($\chi_2^-, \rho_3^-$), the corresponding squared scalar mass matrix has the form:
	\be
	M_{c2} = \left(
	\begin{array}{cc}
		\fr{ \la_8 v_\rho^2}{2} - \fr{A}{2 v^2_\chi} &  \fr{ \la_8 v_\rho v_\chi}{2} - \fr{A}{2 v_\rho v_\chi }\\
		\fr{ \la_8 v_\rho v_\chi}{2} - \fr{A}{2 v_\rho v_\chi } &  \fr{ \la_8 v_\chi^2}{2} - \fr{A}{2 v^2_\rho} \\
	\end{array}
	\right) = -\fr{ (A - \la_8 v_\rho^2 v^2_\chi)}{2} \left(
	\begin{array}{cc}
		\fr{1}{v_\chi^2} &  \fr{1}{v_\chi v_\rho}\\
		\fr{1}{v_\chi v_\rho}&  \fr{1}{v_\rho^2} \\
	\end{array}
	\right)\, .
	\label{c6}
	\ee
This matrix has the massless scalar states $G_2^{\pm}$ and the massive one $H^{\pm}_2$ with mass equal to
	\be m^2_{H^{\pm}_2} =-\fr{ (A - \la_8 v_\rho^2 v^2_\chi)}{2} .  \fr{(v^2_\rho +v^2_\chi )}{v^2_\rho v^2_\chi}
	\label{c7}
	\ee
The physical fields are given as
	\be
	\left(
	\begin{array}{c}
		G_2^{\pm} \\
		H^{\pm}_2 \\
	\end{array}
	\right) = \left(
	\begin{array}{cc}
		\cos \theta_1  & - \sin \theta_1 \\
		\sin \theta_1 &  \cos \theta_1   \\
	\end{array}
	\right) \left(  \begin{array}{c}
		\chi_2^{\pm} \\
		\rho_3^{\pm}\\
	\end{array}
	\right)\, ,
	\label{c8}
	\ee
	where
	\be
	\tan  \theta_1  =\fr{v_\rho}{ v_\chi}\, .
	\label{c4}
	\ee
It is worth mentioning that the bilepton massless $G_2^{\pm}$ correspond to the Goldstone boson associated with the longitudinal component of the $Y^{\pm}$ bilepton gauge boson.
	
From (\ref{c7}) it follows
	\be
	\la_8 > \la_\phi \fr{v_\phi v_\eta}{v_\chi v_\rho} \,. 
	\label{kl2}
	\ee

\subsection{Complex neutral scalar sector}	

There are two neutral scalars: one $\chi^0_1  $ with mass
\be
 m^2_{\chi^0_1}=  ( \la_7 v_\eta^2 v^2_\chi - A)   \fr{(v^2_\eta +v^2_\chi )}{v^2_\eta v^2_\chi}\, .
\label{poten9}
\ee
and one massless $\eta^0_3$ which is identified with Goldstone boson eaten by massive $X^0$.
Hence
\be  \eta^0_3 \equiv G_{X^0} \,.
\label{t22}
\ee
From \eq{poten9},  it follows
	\be
	\la_7 v_\eta^2 v^2_\chi > A \,. 
	\label{t3}
	\ee
It is to be noted that in the framework of  3-3-1 model with right-handed neutrinos,  $\chi^0_1$ is bilepton scalar which can play a role of DM \cite{dmb}.

\subsection{$CP$-odd scalar sector}
	\label{potentialcd}
	
There are four $CP$-odd scalars 
with VEVs:  $(I_\phi, I_{\chi}^3, I_\rho^2 $, $ I_\eta^1)$.
	In the following we describe the corrections to Ref. \cite{jpf}.

	\ben
	\item The squared mass matrix for the electrically neutral $CP$ odd scalars in the basis    {$(I_\phi, I_{\chi}^3, I_\rho , I_\eta^1)$} has the form:
	\bea
	{M_{odd}^2= - \fr A 2  \left(
		\begin{array}{cccc}
			\fr 1 {v_\phi^2} & \fr 1 {v_\phi v_{\chi}}&  \fr 1 {v_\phi v_\rho} &  \fr 1 {v_\phi v_\eta}  \\
			& \fr 1 {v_{\chi}^2}&  \fr 1 {v_{\chi} v_\rho} &  \fr 1 {v_{\chi} v_\eta}  \\
			&  & \fr 1 {v_\rho^2} &  \fr 1 {v_\eta v_\rho} \\
			&  &  & \fr 1 {v_\eta^2} \\
		\end{array}
		\right)\, .
		\label{Modd5}}
	\eea

As seen from Eq.(\ref{Modd5}), there are nontrivial mixings among the $CP$ odd scalars    {$(I_\phi, I_{\chi}^3, I_\rho , I_\eta^1)$} in the interaction basis. Note that an element at the first row and third columns in (\ref{Modd5}) have to be $\fr{1}{v_\rho v_\phi}$, instead of $\fr{1}{v_\rho v_\eta}$ reported in Eq.(16) of Ref.\cite{jpf}.

\item 	The $CP$ odd squared mass matrix $M_{odd}^2$ in (\ref{Modd5}) can be exactly diagonalized by the Euler diagonalization method.
The $CP$ odd scalar fields in the physical and interaction basis are related through the following transformation: 
	{\begin{widetext}
			\be
			\left(
			\begin{array}{c}
				a\\
				G_{Z^\prime} \\
				G_Z	\\
			A_5	 \\
			\end{array}
			\right)  =  
			\left(
			\begin{array}{cccc}
				\cos \theta_\phi & -\sin \theta_3 \sin \theta_\phi &- \sin
				\al \cos \theta_3 \sin \theta_\phi & -\cos \al
				\cos \theta_3 \sin \theta_\phi \\
				0 & \cos \theta_3 & -\sin \al \sin \theta_3 & -\cos
				\al \sin \theta_3 \\
				0 & 0 & \cos \al & -\sin \al \\
				\sin \theta_\phi & \sin \theta_3 \cos \theta_\phi & \sin
				\al \cos \theta_3 \cos \theta_\phi & \cos \al
				\cos \theta_3 \cos \theta_\phi \\
			\end{array}
			\right)
			\left(
			\begin{array}{c}
				I_\phi\\
				I_{\chi}^3 \\
				I_\rho	\\
				I_\eta^1		 \\
				
			\end{array}
			\right)\,,
			\label{physstates}
			\ee
	\end{widetext}}
where the mixing angles in the $CP$ odd scalar sector take the forms:
	{\bea
		&& \tan \al = \fr{v_\eta}{v_\rho}\,, \hs \tan \theta_3 = \fr{v_\eta}{v_{\chi} \sqrt{1+\fr{v_\eta^2}{v_\rho^2}}}\approx \fr{v_\eta}{v_{\chi}}\,, \nn\\
		&& \tan \theta_\phi=\fr{v_{\chi}}{v_\phi \sqrt{1+v_{\chi}^2 \left( \fr{1}{v_\rho^2} + \fr{1}{v_\eta^2}\right)}}\approx \fr{v_{\chi}}{v_\phi}\,. \label{mixodd}
\label{oct3}	
		\eea}

Note that the matrix in \eq{Modd5} depends on four VEVs namely, $v_\rho, v_\eta, v_\chi$ and $v_\phi$. The derived mixing matrix in \eq{physstates} has three angles $\al, \theta_3, \theta_\phi$ given in (\ref{oct3}) and one parameter is $ \left( \fr 1{v^2_\phi} + \fr 1{v^2_{\chi}} + \fr 1{v^2_\rho} +\fr 1{v^2_\eta}  \right)$ which is entered to  expression of $A_5$ mass  in \eq{mA5}. It is worth mentioning that the rotation matrix that diagonalizes the $CP$ odd squared mass matrix has three mixing angles instead of four because of the VEV hierarchy $v_\rho, v_\eta \ll v_\chi \ll v_\phi$.

It is worth mentioning that our result is completely different from the ones given in  Ref.  \cite{a2}, where the mixing matrix is  not unitary.
	
Here the ALP is massless and is given by the following combination of  four $CP$ odd neutral scalar fields  $I_\phi$, $I_{\chi}^3$, $I_\rho$ and $I_\eta^1$:
	\be a = I_\phi \cos \theta_\phi - I_{\chi}^3 \sin \theta_\phi \sin \theta_3 - I_\rho \cos \theta_3 \sin \al \sin \theta_\phi - I_\eta^1 \cos \al \cos \theta_3 \sin \theta_\phi,
		\ee
which  cannot be the same expression for an $a$ given in Refs.\cite{a2,jpf}\footnote{It is possible to get the mixing matrix in which ALP contains only two components as in Refs.\cite{a2,jpf}, but in this case both Goldstone bosons $G_Z $ and $G_{Z^\prime}$	 contain a component along $I_\phi$.}. It is worth mentioning that due to
$v_{\chi} \ll v_\phi$, it follows that	$\tan \theta_\phi \rightarrow 0$ as well as $\sin \theta_\phi$ then $\cos \theta_\phi \backsimeq 1$. This leads to $a \backsimeq I_\phi$.
	
Furthermore, the mass of new massive field $CP$ odd scalar field ${A_5}$ is given by
	\bea m^2_{A_5}  & = &  -\fr{A}{2}  \left( \fr 1{v^2_\phi} + \fr 1{v^2_{\chi}} + \fr 1{v^2_\rho} +\fr 1{v^2_\eta}  \right)\,
	\approx  - \fr 1 2 \la_\phi v_\phi v_{\chi}  \left( \tan \al + \cot \al \right) = -\fr{\la_\phi v_\phi v_\chi}{\sin 2\al}
	\, .\label{mA5}
	\eea
From (\ref{mA5}), we can see that the value of $\la_\phi$ should be negative. It is emphasized that the squared  mass matrix in Eq. \eq{Modd5} as well as mass of the $A_5$ are only  available  due to the last term in \eq{poten3} which just  appears because of	specific discrete symmetry in this paper	(for discussion on this, the reader is referred to Ref. \cite{a2}).
	
\een

Summary: in  the $CP$-odd sector we have  6 fields: two Goldstone bosons for $Z$ and $Z'$, one axion like particle $a$, one massless	 field $G_1$ being eaten by one component of the massive 	 $X^0$   and one massive pseudoscalar $A_5$.

	\subsection{$CP$-even scalar sector}
	\label{potentialce}
	
As same as the $CP$-odd scalar sector, there are four fields in the $CP$-even scalar sector 
with VEVs: $(R_\phi, R_{\chi}^3, R_\rho^2 $ and $ R_\eta^1)$.

	In basis $(R^1_\eta, R_\rho, R_\chi^3, R_\phi)$, the squared mass matrix of $CP$-even  has form as below:
	\bea
	M_R^2 =       {2} \left(
	\begin{array}{cccc}
		\la_2 v_\eta^2-\fr{A}{4 v_\eta^2} &
		\fr{1}{2} \left(\la _6 v_\eta v_\rho+\fr{\la_\phi  v_\chi v_\phi}{2}\right) & \fr{1}{2} \left(\la_4 v_\eta
		v_\chi+\fr{\la_\phi  v_\rho
			v_\phi}{2}\right) & \fr{1}{2} \left(\la _{13}
		v_\eta v_\phi+\fr{\la_\phi
			v_\rho v_\chi}{2}\right) \\
		\fr{1}{2} \left(\la _6 v_\eta v_\rho+\fr{\la_\phi  v_\chi v_\phi}{2}\right) & \la_3 v_\rho^2-\fr{A}{4
			v_\rho^2} & \fr{1}{2} \left(\fr{\la_\phi
			v_\eta v_\phi}{2}+\la _5 v_\rho v_\chi\right) & \fr{1}{2} \left(\fr{\la
			\phi  v_\eta v_\chi}{2}+\lambda _{12}
		v_\rho v_\phi\right) \\
		\fr{1}{2} \left(\la_4 v_\eta v_\chi+\fr{\la_\phi  v_\rho v_\phi}{2}\right) & \fr{1}{2} \left(\fr{\la_\phi
			v_\eta v_\phi}{2}+\la_5 v_\rho v_\chi\right) & \la_1 v_\chi^2-\fr{A}{4 v_\chi^2} & \fr{1}{2}
		\left(\fr{\la\phi  v_\eta v_\rho}{2}+\la_{11} v_\chi v_\phi\right)
		\\
		\fr{1}{2} \left(\la_{13} v_\eta v_\phi+\fr{\la_\phi  v_\rho v_\chi}{2}\right) & \fr{1}{2} \left(\fr{\la_\phi
			v_\eta v_\chi}{2}+\la_{12}
		v_\rho v_\phi\right) & \fr{1}{2}
		\left(\fr{\la_\phi  v_\eta v_\rho}{2}+\la_{11} v_\chi v_\phi\right)
		& \la_{10} v_\phi^2-\fr{A}{4 v_\phi^2} \\
	\end{array}
	\right)\,. \label{MR}
	\eea

Comparing with a similar matrix in Ref.\cite{jpf}, we see that the first three elements in the fourth column of $CP$ even mass matrix in Ref.\cite{a2} have the {\it extra} terms: $\frac{\la_{11} v_\phi v_{\chi^{\prime}}}{2},\fr{\la_{13} v_\phi v_\eta}{2} $ and $\fr{\la_{12} v_\phi v_\rho}{2}$, respectively. To recognize the existence of these terms, let us write them explicitly
\bea
&& \la_{11} ( \phi^\dag \phi)( \chi^\dag \chi)  \supset v_\phi v_{\chi^{\prime}} R_\phi R_{\chi^{\prime}} , \crn
&& \la_{12} ( \phi^\dag \phi)( \rho^\dag \rho) \supset  v_\phi v_\rho R_\phi R_\rho ,\crn
&& \la_{13} ( \phi^\dag \phi)( \eta^\dag \eta) \supset  v_\phi v_\eta R_\phi R_\eta .\nn
\eea

The matrix which is used to diagonalize $M_R^2$ is:
	\bea U_R 
	= \left(
	\begin{array}{cccc}
		-\cos \al_2 & -\sin \al_2 \cos \al_3 & -\sin \al_2 \sin \al_3 \cos \al_\phi & \sin \al_2 \sin \al_3 \sin \al_\phi \\
		\sin \al_2 & -\cos \al_2 \cos \al_3 & -\cos \al_2 \sin \al_3 \cos \al_\phi & \cos \al_2 \sin \al_3 \sin \al_\phi \\
		0 & \sin \al_3 & -\cos \al_3 \cos \al_\phi & \cos \al_3 \sin \al_\phi \\
		0 & 0 & \sin \al_\phi & \cos \al_\phi \\
	\end{array}
	\right)\,
	\eea
in which, the mixing angles in the $CP$ even scalar sector are defined as below:
	\bea
	&& \tan 2 \al_2= \fr{4 \cos \al_3 v_\eta v_\rho (A + \la_6 v_\eta^2 v_\rho^2)}{A \cos^2 \al_3 v_\eta^2 -A v_\rho^2 + 4 v_\eta^2 v_\rho^2 (\la_2 v_\eta^2 - \la_3 \cos^2 \al_3 v_\rho^2)}\, \\
	&&  \tan 2\al_3 = \fr{4 v_\chi
		\left(A+2 \la_5 v_\rho^2
		v_\chi^2\right)}{ \cos \alpha
			\phi  \left(A-4 \la_1	v_\chi^4\right)^2}\,, \\
	&& \tan 2 \al_\phi = \fr{\la_{11} v_{\chi }}{\la_{10} v_{\phi }} \,.
	\eea

Changing the signs of $h$ , $h_5$ and $H_\chi$, the physical fields are given by:
		\begin{widetext}
			\be
			\left(
			\begin{array}{c}
				h_5\\
				h \\
				H_\chi	\\
				\Phi	 \\
			\end{array}
			\right)  =   \left(
			\begin{array}{cccc}
				\cos \al_2 & \sin \al_2 \cos \al_3 & \sin \al_2 \sin \al_3 \cos \al_\phi & -\sin \al_2 \sin \al_3 \sin \al_\phi \\
				-\sin \al_2 & \cos \al_2 \cos \al_3 & \cos \al_2 \sin \al_3 \cos \al_\phi & -\cos \al_2 \sin \al_3 \sin \al_\phi \\
				0 & -\sin \al_3 & \cos \al_3 \cos \al_\phi & -\cos \al_3 \sin \al_\phi \\
				0 & 0 & \sin \al_\phi & \cos \al_\phi \\
			\end{array}
			\right)
			\left(
			\begin{array}{c}
				R_\eta^{1}\\
				R_\rho \\
				R_\chi^{3}	\\
				R_\phi	 \\
			\end{array}
			\right)\,.
			\label{physstateR}
			\ee
		\end{widetext}

In the limit $v_\phi \gg v_\chi \gg v_\rho, v_\eta$ it follows
\bea
		h_5 &\approx& R^1_\eta \cos \al_2+ R_\rho \sin \al_2 \,,\\
		h &\approx& -R^1_\eta \sin \al_2+ R_\rho\cos \al_2  \,,\\
		H_\chi &\approx& R^3_\chi   \cos \al_\phi\,,\\
		\Phi &\approx& R_\phi \cos \al_\phi \,,\label{physstateg}
		\eea	
and their respective masses are shown in Appendix.(\ref{$CP$e}).	

Note that comparing to the $4 \times 4$ matrix of $CP$-odd sector containing only four parameters
	with three massless solution,  the matrix in \eq{MR} having 10 parameters is not exactly diagonalized. To solve this problem we have used the Hatree-Fock method where some conditions
	such as $v_\phi \gg v_\chi \gg v_\rho, v_\eta$, $\la_\phi \ll 1$  and $\sin \al_3 \approx 0$.
	As a consequence of the aforementioned VEV hierarchy, the derived matrix contains three angles $\al_2,\al_3$ and $\al_\phi$ and three
	parameters associated  with masses of new fields $\Phi, H_\chi$ and $h_5$.

In the limit $v_\phi \gg v_\chi \gg v_\rho \gg v_\eta $, one has
		\bea
		\chi  &\simeq &
		\left(
		\begin{array}{c}
			\chi_1^0 \\
			G_{Y^-} \\
			\fr 1{\sqrt{2}}\left( v_\chi  + H_\chi + i G_{Z'}\right) \\
		\end{array}
		\right),\,
		\eta \simeq \left(
		\begin{array}{c}
			\fr{1}{\sqrt{2}}\left(  u +h_5 + i A_5\right) \\
			H_1^- \\
			G_{X^0}   \\
		\end{array}
		\right),\,
		\rho \simeq 
		\left(
		\begin{array}{c}
			G_{W^+} \\
			\fr{1}{\sqrt{2}}\left(  v + h+ i G_Z\right)\\
			H_2^+ \\
		\end{array}
		\right)\,,
		\crn
		\phi & \simeq & \fr{1}{\sqrt{2}}\left( v_\phi + \Phi + i a \right)\, .
		\label{d25}\eea
	
In the $CP$-even scalar sector, there are six fields. One massless field is part of $G_{X^0}$, another massive in TeV scale is associated to $\chi_1^0$. One heavy field with mass in the range of $10^{11}$ GeV and associated with singlet $\phi$ is identified  to inflaton $\Phi$. One SM-like Higgs boson $h$ with mass $\sim$ 125 GeV. Two remain fields include one heavy with mass at TeV scale ($H_\chi$) and another with mass at EW scale ($h_5$).
	
Combination of table \ref{qnumber} and \eq{d25} leads to some interesting consequences
\ben

\item SM-like Higgs boson $h$  has Yukawa couplings with only  SM fermions

\item ALP $a$ can have Yukawa couplings with only exotic quarks.

\item The pseudoscalar  $A_5$ and $H_\chi$ can have Yukawa couplings with not only exotic quarks but also SM quarks and leptons.

\een

\section{Numerical analysis of the scalar sector}
\label{numer}

To find particle content in $CP$-even sector namely the SM-like Higgs boson, and another 
one close to it $H_5$ is the aim in this section.

\ben
	\item	In order to successfully reproduce the $W$ gauge boson mass, the VEVs of the $SU(3)_L$ scalar triplets $\eta$ and $\rho$ should obey the following constraint:
	\be
	v_\eta = \sqrt{v^2 - v_\rho^2}\,.
	\label{j291}
	\ee
	where $v=246$ GeV is the electroweak symmetry breaking scale.

	\item Charged sector
	
	a) From Eq. (\eq{c2}), it follows that positive squared scalar masses are obtained provided that the following relation is fulfilled:
	
	\be   \la_9 v_\rho^2 v^2_\eta \,  > \, A
	\label{j292}
	\ee
	
	b) From  Eq.  \eq {c7}, it follows that \be  \la_8 v_\rho^2 v^2_\chi  \, > \, A
	\label{j293}
	\ee
	
	\item $CP$ -odd sector
	
	i) From Eq. \eq{poten9} it follows that the requirement of obtaining positive squared mass for the  massive complex scalar $\va^0$ implies:
	\be
	\la_7 v_\eta^2 v^2_\chi \, > \, A \, .
	\label{j294}
	\ee
	
	ii) From Eq. \eq{mA5}, it follows   \bea m^2_{A_5}  & = & -\fr{A}{2}  \left( \fr 1{v^2_\phi} + \fr 1{v^2_{\chi}} + \fr 1{v^2_\rho} +\fr 1{v^2_\eta}  \right)\,
\simeq  -\frac{\la_\phi v_\phi v_\chi}{\sin 2\al}
	\, .\label{A295}
	\eea
	 If $v_\eta = v_\rho$ in EW scale, then we may have $\left(m_{A_5}^2\right)_{min} = -\la_\phi v_\phi v_\chi $, which implies $\la_\phi<0$.
	 From Eq.(\eqref{A295}), we get $\la_\phi = - \fr{m_{A_5}^2 \sin 2\al}{v_\phi v_\chi}$. With $m_{A_5} \sim 10^3$ GeV, $v_\phi \sim 10^{10}$ GeV and $v_\chi = 10^5$ GeV, then we get  $\vert \la_\phi \vert < 10^{-9}$.
	 Moreover, from the condition for $\la_9$ and  assuming $v_\eta = v_\rho \simeq 174$ GeV, $v_\phi = 10^{10}$ GeV and $v_\chi = 10^5$ GeV, then we get  $\vert \la_\phi \vert < 10^{-10}$.
	 The tiny value of the quartic scalar coupling $\lambda_\phi$ can be qualitatively understood from the requirement of having a physical pseudoscalar $A_5$ with a mass at the TeV or subTeV scale. It is worth mentioning that the $Z_{11}$ symmetry is spontaneously broken at a very large scale $\sim  10^{10}$ GeV by the VEV 
	 	of the singlet scalar field $\phi$, which also generates the mass for the physical pseudoscalar $A_5$. Another more formal way to justify the smallness of $\lambda_\phi$ is by considering an accidental Peccei - Quinn symmetry $U(1)_{PQ}$ under which $\phi$ has charge equal to $-2$, whereas the right handed Majorana neutrinos, the $SU(3)_L$ leptonic triplets and the right handed leptons will have charges equal to $1$. Under that assignment the quartic scalar interaction involving $\lambda_\phi$ will be forbidden at tree level, however the mass of the pseudoscalar $A_5$ can be radiatively generated from a box diagram involving the one loop level exchange of the neutral components of the $SU(3)_L$ scalar triplets 
	 	as well as the exchange of the scalar singlet $\phi$. That loop suppression together with the large mass scale of the $CP$ even component of $\phi$ can be interpreted as dynamical sources for the tiny values of the $\la_\phi$ coupling. Besides that, it is worth mentioning that low energy effective theory below the scale of breaking of the $SU(3)_L\times U(1)_X$ gauge symmetry corresponds to a Two Higgs Doublet Model, where the consistency with allowed experimental ranges for the oblique $T$, $S$ and $U$ parameters, requires that the masses of the non SM scalars should not differ significantly \cite{CarcamoHernandez:2015smi}. In view of the above, it is required that the pseudoscalar $A_5$ should acquire a mass at the subTeV or TeV scale, not far from the masses of the physical scalar states arising from the $\eta$ and $\rho$ scalar triplets.

	\item $CP$-even sector
	\bit
	\item  Mass of inflaton
	\be m_\Phi = \sqrt{2  \la_{10}}\,  v_\phi  \, \approx \,  10^{11} \, \textrm{GeV} \hs  \Rightarrow  \la_{10} \approx  1\,\hs  \textrm{if}\hs  \,  v_\phi \approx 10^{10}  \, \textrm{GeV}\,.
	\ee
		
	\item  Mass of heavy scalar: The Eq.\eq{b1} yields
	\be m_{H_{\chi}}^2   {\approx 2 \la_1 v_\chi^2 + \fr{\la_5^2}{2 \la_1} v_\rho^2}\,.
	\label{A296}	\ee
		
	\item Two light scalars: From the Eq.(\ref{mhh5}) and use the approximation $\la_2 \simeq \la_3 \simeq \la_6$ we have:
		\bea
		m_{h,h5}^2 &\approx&\la_3 v^2+\fr{m_{A_5}^2}{2 } \pm \sqrt{m_{A_5}^4 +\la_3^2 \left(v^4-3v_\eta^2 v_\rho^2\right)-\fr{\la_3 m_{A_5}^2
				\left(v^4-2v_\eta^2 v_\rho^2\right)}{v^2}}\,.
		\eea

In case $v_\eta = v_\rho = \fr{v}{\sqrt2}$, the model predicts
\bea
m_{h,h5}^2 &\simeq&  
\lambda _3 v^2 +\fr{m_{A_5}^2}{2 } \pm \fr{\la_3 v^2-m_{A_5}^2}{2} \,.
\eea

Then we have:
\bea
m_h^2 && \simeq \fr{3}{2} \la_3 v^2
\,,\label{hSM}\\
m_{h_5}^2 && \simeq \fr{\la_3 v^2}{2}  +m_{A_5}^2 \label{h5}
\eea

\eit
\een

One scalar is the SM like Higgs boson $h$ with mass of 125 GeV. One another scalar is a new one $h_5$ with mass takes the values of $150$ GeV \cite{7a,8a,9a,10a,11a,12a,13a} or $96$  GeV  \cite{96gev1,96gev2,96gev3,96gev4}, respectively. The mass value of $h_5$ depends on some parameters such as $ \la_2, \la_3, \la_\phi$ and the VEVs of the scalar fields in this model. 
From  \eq{hSM} and \eq{h5}, we have the correlation between $A_5, h$ and $h_5$ as below:
\bea
|m_{h_5}^2-m_{A_5}^2| = \mathcal{O}\left(m_h^2\right)\,.\label{ch5A5}
\eea
From Eq.\eq{ch5A5}, it follows that in the case $v_\eta = v_\rho$, the splitting by masses of $h_5$ and $A_5$ is about few hundreds GeV.

\section{Yukawa couplings and top quark FCNC decays}
In the quark sector, there are two parts: exotic quarks without mass mixing and ordinary quarks with mass mixing. Because of having no mass mixing, the mass eigenstates of exotic quarks are their original states. Then, we just consider on the mass mixing of ordinary quarks. The mass matrices of ordinary quarks are
\be 
	M_{u}=\left( 
	\begin{array}{ccc}
		\left( y_{6}\right) _{11}\fr{v_{\rho }}{v_{\eta }} & \left( y_{6}\right)
		_{12}\frac{v_{\rho }}{v_{\eta }} & \left( y_{6}\right) _{13}\fr{v_{\rho }}{%
			v_{\eta }} \\ 
		\left( y_{6}\right) _{21}\fr{v_{\rho }}{v_{\eta }} & \left( y_{6}\right)
		_{22}\fr{v_{\rho }}{v_{\eta }} & \left( y_{6}\right) _{23}\fr{v_{\rho }}{%
			v_{\eta }} \\ 
		\left( y_{3}\right) _{31} & \left( y_{3}\right) _{32} & \left( y_{3}\right)
		_{33}%
	\end{array}%
	\right) \fr{v_{\eta }}{\sqrt{2}}=V_{uL}\widetilde{M}_{u}V_{uR}^{\dagger }\,, \label{Massu}
\end{equation}
with 
\be 
	\widetilde{M}_{u}=diag\left( m_{u},m_{c},m_{t}\right)\,
\ee 
and 
\be 
	M_{d}=\left( 
	\begin{array}{ccc}
		\left( y_{4}\right) _{11}\fr{v_{\eta }}{v_{\rho }} & \left( y_{4}\right)
		_{12}\fr{v_{\eta }}{v_{\rho }} & \left( y_{4}\right) _{13}\fr{v_{\eta }}{%
			v_{\rho }} \\ 
		\left( y_{4}\right) _{21}\frac{v_{\eta }}{v_{\rho }} & \left( y_{4}\right)
		_{22}\fr{v_{\eta }}{v_{\rho }} & \left( y_{4}\right) _{23}\fr{v_{\eta }}{%
			v_{\rho }} \\ 
		\left( y_{5}\right) _{31} & \left( y_{5}\right) _{32} & \left( y_{5}\right)
		_{33}%
	\end{array}%
	\right) \fr{v_{\rho }}{\sqrt{2}}=V_{dL}\widetilde{M}_{d}V_{dR}^{\dagger } \,, \label{Massd}
\ee
with
\be 
	\widetilde{M}_{d}=diag\left( m_{d},m_{s},m_{b}\right)  \,.
\ee 
\be 
	K=V_{uL}^{\dagger }V_{dL}  \,.
\ee 

In these matrices above, all Yukawa couplings of the form $\left(y_i\right)_{ab}$ $a,b=1,2,3$; $i=3,4,5,6$ are real and positive. With $\al =1, 2$ and $a=\al, 3$, these couplings can be defined by the following equations:
\be 
	\left( y_{6}\right) _{n a}=\fr{\sqrt{2}}{v_{\rho }}\left( V_{uL}\widetilde{M%
	}_{u}V_{uR}^{\dagger }\right) _{n a},\hspace{1cm}\hspace{1cm}\left(
	y_{3}\right) _{3a}=\fr{\sqrt{2}}{v_{\eta }}\left( V_{uL}\widetilde{M}%
	_{u}V_{uR}^{\dagger }\right) _{3a}  \label{yu}
\end{equation}
\be 
	\left( y_{4}\right) _{n a}=\fr{\sqrt{2}}{v_{\eta }}\left( V_{dL}\widetilde{M%
	}_{d}V_{dR}^{\dagger }\right) _{n a},\hspace{1cm}\hspace{1cm}\left(
	y_{5}\right) _{3a}=\fr{\sqrt{2}}{v_{\rho }}\left( V_{dL}\widetilde{M}%
	_{d}V_{dR}^{\dagger }\right) _{3a}\,.  \label{yd}
\ee 
  From (\ref{Massu}) and (\ref{Massd}), the diagonalized mass matrix of ordinary quarks are defined as below:
\be 
	\widetilde{M}_{u,d}=\left( V_{L}^{\left( u,d\right) }\right) ^{\dagger
	}M_{u,d}V_{R}^{\left( u,d\right) }\,.  \label{MSMquarkdiag}
\ee
In general, we get:
\bea 
	\widetilde{M}_{f} &=&\left( M_{f}\right) _{diag}=V_{fL}^{\dagger
	}M_{f}V_{fR},\hspace{1cm}\hspace{1cm}f_{\left( L,R\right) }=V_{f\left(
		L,R\right) }\widetilde{f}_{\left( L,R\right) },  \crn 
	\overline{f}_{aL}\left( M_{f}\right) _{ab}f_{bR} &=&\overline{\widetilde{f}}%
	_{kL}\left( V_{fL}^{\dagger }\right) _{ka}\left( M_{f}\right) _{ab}\left(
	V_{fR}\right) _{bl}\widetilde{f}_{lR}=\overline{\widetilde{f}}_{kL}\left(
	V_{fL}^{\dagger }M_{f}V_{fR}\right) _{kl}\widetilde{f}_{lR}=\overline{%
		\widetilde{f}}_{kL}\left( \widetilde{M}_{f}\right) _{kl}\widetilde{f}%
	_{lR}=m_{f_k}\overline{\widetilde{f}}_{kL}\widetilde{f}_{kR}, \crn 
	k &=&1,2,3\,. \label{physfermions}
\eea %
Here, $\widetilde{f}_{k\left( L,R\right) }$ and $f_{k\left( L,R\right) }$ ($%
k=1,2,3$) are the SM fermionic fields in the mass and interaction bases,
respectively. Hence, the SM up and down type quark Yukawa interactions are given by:
\bea 
-\mathcal{L}_{Y}^{\left( u\right) } &=&\sum_{n=1}^{2}\sum_{a=1}^{3}\left(
y_{6}\right) _{na}\overline{u}_{nL}\fr{v_{\rho }+R_{\rho }-iI_{\rho }}{%
	\sqrt{2}}u_{aR}+\sum_{a=1}^{3}\left( y_{3}\right) _{3a}\overline{u}_{3L}%
\frac{v_{\eta }+R_{\eta }^{1}+iI_{\eta }^{1}}{\sqrt{2}}u_{bR}+h.c  \crn 
&=&\sum_{n=1}^{2}\sum_{a=1}^{3}\sum_{b=1}^{3}\sum_{c=1}^{3}\left(
y_{6}\right) _{na}\overline{\widetilde{u}}_{cL}\left( \left( V_{L}^{\left(
	u\right) }\right) ^{\dagger }\right) _{cn}\fr{v_{\rho }+R_{\rho }-iI_{\rho
}}{\sqrt{2}}\left( V_{R}^{\left( u\right) }\right) _{ab}\widetilde{u}_{bR} 
\crn 
&&+\sum_{a=1}^{3}\sum_{b=1}^{3}\sum_{c=1}^{3}\left( y_{3}\right) _{3a}%
\overline{\widetilde{u}}_{cL}\left( \left( V_{L}^{\left( u\right) }\right)
^{\dagger }\right) _{c3}\fr{v_{\eta }+R_{\eta }^{1}+iI_{\eta }^{1}}{\sqrt{2%
}}\left( V_{R}^{\left( u\right) }\right) _{ab}\widetilde{u}_{bR}+h.c  \crn 
&=&\sum_{n=1}^{2}\sum_{a=1}^{3}\sum_{b=1}^{3}\sum_{c=1}^{3}\fr{\sqrt{2}}{%
	v_{\rho }}\left( V_{uL}\widetilde{M}_{u}V_{uR}^{\dagger }\right) _{na}%
\overline{\widetilde{u}}_{cL}\left( \left( V_{L}^{\left( u\right) }\right)
^{\dagger }\right) _{cn}\fr{v_{\rho }+R_{\rho }-iI_{\rho }}{\sqrt{2}}%
\left( V_{R}^{\left( u\right) }\right) _{ab}\widetilde{u}_{bR}  \crn 
&&+\sum_{a=1}^{3}\fr{\sqrt{2}}{v_{\eta }}\left( V_{uL}\widetilde{M}%
_{u}V_{uR}^{\dagger }\right) _{3a}\overline{\widetilde{u}}_{cL}\left( \left(
V_{L}^{\left( u\right) }\right) ^{\dagger }\right) _{c3}\frac{v_{\eta
	}+R_{\eta }^{1}+iI_{\eta }^{1}}{\sqrt{2}}\left( V_{R}^{\left( u\right)
}\right) _{ab}\widetilde{u}_{bR}+h.c\,.  \label{LYu}
\eea 
\bea 
-\mathcal{L}_{Y}^{\left( d\right) } &=&\sum_{n=1}^{2}\sum_{a=1}^{3}\left(
y_{4}\right) _{na}\overline{d}_{nL}\frac{v_{\eta }+R_{\eta }^{1}-iI_{\eta
	}^{1}}{\sqrt{2}}d_{aR}+\sum_{a=1}^{3}\left( y_{5}\right) _{3a}\overline{d}%
_{3L}\frac{v_{\rho }+R_{\rho }+iI_{\rho }}{\sqrt{2}}d_{bR}+h.c  \crn 
&=&\sum_{n=1}^{2}\sum_{a=1}^{3}\sum_{b=1}^{3}\sum_{c=1}^{3}\left(
y_{4}\right) _{na}\overline{\widetilde{d}}_{cL}\left( \left( V_{L}^{\left(
	d\right) }\right) ^{\dagger }\right) _{cn}\fr{v_{\eta }+R_{\eta
	}^{1}-iI_{\eta }^{1}}{\sqrt{2}}\left( V_{R}^{\left( d\right) }\right) _{ab}%
\widetilde{d}_{bR}  \crn 
&&+\sum_{a=1}^{3}\sum_{b=1}^{3}\sum_{c=1}^{3}\left( y_{5}\right) _{3a}%
\overline{\widetilde{d}}_{cL}\left( \left( V_{L}^{\left( d\right) }\right)
^{\dagger }\right) _{c3}\fr{v_{\rho }+R_{\rho }+iI_{\rho }}{\sqrt{2}}%
\left( V_{R}^{\left( d\right) }\right) _{ab}\widetilde{d}_{bR}+h.c  \crn 
&=&\sum_{n=1}^{2}\sum_{a=1}^{3}\sum_{b=1}^{3}\sum_{c=1}^{3}\frac{\sqrt{2}}{%
	v_{\eta }}\left( V_{dL}\widetilde{M}_{d}V_{dR}^{\dagger }\right) _{na}%
\overline{\widetilde{d}}_{cL}\left( \left( V_{L}^{\left( d\right) }\right)
^{\dagger }\right) _{cn}\fr{v_{\eta }+R_{\eta }^{1}-iI_{\eta }^{1}}{\sqrt{2%
}}\left( V_{R}^{\left( d\right) }\right) _{ab}\widetilde{d}_{bR}  \crn 
&&+\sum_{a=1}^{3}\sum_{b=1}^{3}\sum_{c=1}^{3}\frac{\sqrt{2}}{v_{\rho }}%
\left( V_{dL}\widetilde{M}_{d}V_{dR}^{\dagger }\right) _{3a}\overline{%
	\widetilde{d}}_{cL}\left( \left( V_{L}^{\left( d\right) }\right) ^{\dagger
}\right) _{c3}\fr{v_{\rho }+R_{\rho }+iI_{\rho }}{\sqrt{2}}\left(
V_{R}^{\left( d\right) }\right) _{ab}\widetilde{d}_{bR}+h.c\,.
\label{LYd}
\eea 
Replacing Eqs. (\ref{physstates}) and (\ref{physstateR}) in (\ref{LYu}) and (\ref{LYd}), we found that the Yukawa couplings of $h$, $h_5$ and $A_5$ with up and down -type SM quarks are given by:
\bea 
\left( \Ga_{u}^{h}\right) _{ij} &=&\fr{\cos \alpha _{2}}{v_{\rho }}%
\sum_{n=1}^{2}\sum_{a=1}^{3}\left( \left( V_{L}^{\left( u\right) }\right)
^{\dagger }\right) _{in}\left( V_{uL}\widetilde{M}_{u}V_{uR}^{\dagger
}\right) _{na}\left( V_{R}^{\left( u\right) }\right) _{aj}  \notag \\
&&-\fr{\sin \alpha _{2}}{v_{\eta }}\sum_{a=1}^{3}\left( \left(
V_{L}^{\left( u\right) }\right) ^{\dagger }\right) _{i3}\left( V_{uL}%
\widetilde{M}_{u}V_{uR}^{\dagger }\right) _{3a}\left( V_{R}^{\left( u\right)
}\right) _{aj} \label{ghu}
\eea 
\bea 
\left( \Ga_{u}^{h_{5}}\right) _{ij} &=&\fr{\sin \alpha _{2}}{v_{\rho }}%
\sum_{n=1}^{2}\sum_{a=1}^{3}\left( \left( V_{L}^{\left( u\right) }\right)
^{\dagger }\right) _{in}\left( V_{uL}\widetilde{M}_{u}V_{uR}^{\dagger
}\right) _{na}\left( V_{R}^{\left( u\right) }\right) _{aj}  \crn 
&&+\fr{\cos \alpha _{2}}{v_{\eta }}\sum_{a=1}^{3}\left( \left(
V_{L}^{\left( u\right) }\right) ^{\dagger }\right) _{i3}\left( V_{uL}%
\widetilde{M}_{u}V_{uR}^{\dagger }\right) _{3a}\left( V_{R}^{\left( u\right)
}\right) _{aj}
\eea 
\bea 
\left( \Ga_{u}^{A_{5}}\right) _{ij} &=&-i\fr{\sin \alpha }{v_{\rho }}%
\sum_{n=1}^{2}\sum_{a=1}^{3}\left( \left( V_{L}^{\left( u\right) }\right)
^{\dagger }\right) _{in}\left( V_{uL}\widetilde{M}_{u}V_{uR}^{\dagger
}\right) _{na}\left( V_{R}^{\left( u\right) }\right) _{aj}  \crn 
&&+i\fr{\cos \alpha }{v_{\eta }}\sum_{a=1}^{3}\left( \left( V_{L}^{\left(
	u\right) }\right) ^{\dagger }\right) _{i3}\left( V_{uL}\widetilde{M}%
_{u}V_{uR}^{\dagger }\right) _{3a}\left( V_{R}^{\left( u\right) }\right)
_{aj}
\eea 
\bea 
\left( \Ga_{d}^{h}\right) _{ij} &=&-\fr{\sin \alpha _{2}}{v_{\eta }}%
\sum_{n=1}^{2}\sum_{a=1}^{3}\left( \left( V_{L}^{\left( d\right) }\right)
^{\dagger }\right) _{in}\left( V_{dL}\widetilde{M}_{d}V_{dR}^{\dagger
}\right) _{na}\left( V_{R}^{\left( d\right) }\right) _{aj}  \crn 
&&+\fr{\cos \alpha _{2}}{v_{\rho }}\sum_{a=1}^{3}\left( \left(
V_{L}^{\left( d\right) }\right) ^{\dagger }\right) _{i3}\left( V_{dL}%
\widetilde{M}_{d}V_{dR}^{\dagger }\right) _{3a}\left( V_{R}^{\left( d\right)
}\right) _{aj} \label{ghd}
\eea 
\bea 
\left( \Ga_{d}^{h_{5}}\right) _{ij} &=&\fr{\cos \alpha _{2}}{v_{\eta }}%
\sum_{n=1}^{2}\sum_{a=1}^{3}\left( \left( V_{L}^{\left( d\right) }\right)
^{\dagger }\right) _{in}\left( V_{dL}\widetilde{M}_{d}V_{dR}^{\dagger
}\right) _{na}\left( V_{R}^{\left( d\right) }\right) _{aj}  \crn 
&&+\frac{\sin \al_{2}}{v_{\rho }}\sum_{a=1}^{3}\left( \left(
V_{L}^{\left( d\right) }\right) ^{\dagger }\right) _{i3}\left( V_{dL}%
\widetilde{M}_{d}V_{dR}^{\dagger }\right) _{3a}\left( V_{R}^{\left( d\right)
}\right) _{aj}
\eea 
\bea 
\left( \Ga_{d}^{A_{5}}\right) _{ij} &=&-i\fr{\cos \alpha }{v_{\eta }}%
\sum_{n=1}^{2}\sum_{a=1}^{3}\left( \left( V_{L}^{\left( d\right) }\right)
^{\dagger }\right) _{in}\left( V_{dL}\widetilde{M}_{d}V_{dR}^{\dagger
}\right) _{na}\left( V_{R}^{\left( d\right) }\right) _{aj}  \crn
&&+i\fr{\sin \alpha }{v_{\rho }}\sum_{a=1}^{3}\left( \left( V_{L}^{\left(
	d\right) }\right) ^{\dagger }\right) _{i3}\left( V_{dL}\widetilde{M}%
_{d}V_{dR}^{\dagger }\right) _{3a}\left( V_{R}^{\left( d\right) }\right)
_{aj}
\eea 

Rewriting the couplings (\ref{ghu}) and  (\ref{ghd}) in another form, one gets:
\bea
\left( \Ga_{u,d}^{h}\right) _{ij} &=& \fr{\cos \al_2}{v_\rho} \left(\widetilde{M}_{u,d}\right)_{ij}-\fr{\cos \al_2}{v_\eta} (\tan \al + \tan \al_2) \left(\Ga_h^{\prime (u,d)}\right)_{ij}\,. \label{ghud}
\eea
The first term in (\ref{ghud}) is a flavor conserving. The second term in (\ref{ghud}) is a flavor changing. In order to have flavor conservation for SM-Higgs interactions, the second term should be vanished. Then, one gets the condition below:
\be
\tan \al = - \tan \al_2 \label{dkaa2}
\ee
The Eq.(\ref{dkaa2}) gives the condition among $v_\rho$ and $v_\phi, v_\chi, \la_\phi, \la_2, \la_3, \la_6$ which guarantees the flavor conservation of SM-Higgs at tree level.
In the SM, the resulting top quark FCNCs are strongly suppressed. But in this model, the FCNCs of top quark appear and can be used to look for new physics. The Yukawa couplings of up-type quarks $\Ga_{ut,ct}^{h,h_5}$ allow some decays at tree-level such as: $t \rightarrow hu$ or $t \rightarrow hc$. These processes get the branching ratios limited by ATLAS \cite{thqATLAS}: at 95\% C.L. upper limits on the $Br(t \rightarrow hc) = 1.1 \times  10^{-3} (8.3 \times 10^{-4} )$ and  $Br(t \rightarrow hu) = 1.2 \times  10^{-3} (8.3 \times 10^{-4} )$, respectively. The corresponding combined observed (expected) upper limits on the couplings $| \Ga_{tc}^h  | =  0.064 (0.055)$  and $| \Ga_{tu}^h  |= 0.066 (0.055)$, respectively. 

Considering the process $t \rightarrow hc$, its branching ratio is given by:
\be
Br(t\rightarrow hc) = \fr{\fr{g_{thc}^2}{4\pi}\fr{(m_t^2-m_h^2)^2}{2m_t m_h}}{\Ga_t}\,,
\ee
with $\Ga_t =1.32 GeV$ is the decay width for top quark ($m_t = 172.5 GeV$) predicted by SM. And $g_{thc}$ is the coupling defined by \cite{ghtc}:
 \bea
 g_{thc}^2 &=& \left(\fr{\cos \alpha _{2}}{v_{\rho }}%
 \left( \left( V_{L}^{\left( u\right) }\right)
 ^{\dagger }\right) _{23}\left( V_{uL}\widetilde{M}_{u}V_{uR}^{\dagger
 }\right) _{32}\left( V_{R}^{\left( u\right) }\right) _{23} -\frac{\sin \alpha _{2}}{v_{\eta }} \left( \left( V_{L}^{\left( u\right) }\right) ^{\dagger }\right) _{23}\left( V_{uL} \widetilde{M}_{u}V_{uR}^{\dagger }\right) _{32}\left( V_{R}^{\left( u\right) }\right) _{23}\right)^2\nn\\
&& + \left(\frac{\cos \alpha _{2}}{v_{\rho }}\left( \left( V_{L}^{\left( u\right) }\right)
^{\dagger }\right) _{32}\left( V_{uL}\widetilde{M}_{u}V_{uR}^{\dagger
}\right) _{23}\left( V_{R}^{\left( u\right) }\right) _{32} -\frac{\sin \alpha _{2}}{v_{\eta }} \left( \left( V_{L}^{\left( u\right) }\right) ^{\dagger }\right) _{32}\left( V_{uL} \widetilde{M}_{u}V_{uR}^{\dagger }\right) _{23}\left( V_{R}^{\left( u\right) }\right) _{32}\right)^2 
 \eea
We can also get the branching ratio for the process $t \rightarrow h u$ as follows:
\be
Br(t\rightarrow hu) = \fr{\fr{g_{thu}^2}{4\pi}\fr{(m_t^2-m_h^2)^2}{2m_t m_h}}{\Ga_t}\,,
\ee
with $g_{thu}$ is the coupling that is similarly defined by:
\bea
g_{thu}^2 &=& \left(\fr{\cos \al_{2}}{v_{\rho }}%
\left( \left( V_{L}^{\left( u\right) }\right)
^{\dagger }\right) _{13}\left( V_{uL}\widetilde{M}_{u}V_{uR}^{\dagger
}\right) _{31}\left( V_{R}^{\left( u\right) }\right) _{13} -\fr{\sin \al_{2}}{v_{\eta }} \left( \left( V_{L}^{\left( u\right) }\right) ^{\dagger }\right) _{13}\left( V_{uL} \widetilde{M}_{u}V_{uR}^{\dagger }\right) _{31}\left( V_{R}^{\left( u\right) }\right) _{13}\right)^2\nn\\
&& + \left(\frac{\cos \al_{2}}{v_{\rho }}\left( \left( V_{L}^{\left( u\right) }\right)
^{\dagger }\right) _{31}\left( V_{uL}\widetilde{M}_{u}V_{uR}^{\dagger
}\right) _{13}\left( V_{R}^{\left( u\right) }\right) _{31} -\fr{\sin \alpha _{2}}{v_{\eta }} \left( \left( V_{L}^{\left( u\right) }\right) ^{\dagger }\right) _{31}\left( V_{uL} \widetilde{M}_{u}V_{uR}^{\dagger }\right) _{13}\left( V_{R}^{\left( u\right) }\right) _{31}\right)^2 
\eea
With $m_h = 125$ GeV, $m_t=172.9$ GeV and the branching ratios limited by ATLAS that we mentioned above, we plot the correlation between the mixing angle in range $(-\fr{\pi}{2},-\fr{\pi}{4})$ and the branching ratios of $t \rightarrow hq$ decay with $q=u,c$.
Moreover, in this model, we have a light non SM $CP$ even scalar field such as $h_5$ then the decays $t \rightarrow q h_5 $ ($q=c, u$) can be under the consideration as well as the decays $t \rightarrow h q$. The couplings of the decays $t \rightarrow q h_5 $ ($q=c, u$) are defined by:
\bea
g_{th_5 q_i}^2 &=& \left(\fr{-\sin \alpha _{2}}{v_{\rho }}%
\left( \left( V_{L}^{\left( u\right) }\right)
^{\dagger }\right) _{i3}\left( V_{uL}\widetilde{M}_{u}V_{uR}^{\dagger
}\right) _{3i}\left( V_{R}^{\left( u\right) }\right) _{i3} -\fr{\cos \alpha _{2}}{v_{\eta }} \left( \left( V_{L}^{\left( u\right) }\right) ^{\dagger }\right) _{i3}\left( V_{uL} \widetilde{M}_{u}V_{uR}^{\dagger }\right) _{3i}\left( V_{R}^{\left( u\right) }\right) _{i3}\right)^2\nn\\
&& + \left(\fr{-\sin \al_{2}}{v_{\rho }}\left( \left( V_{L}^{\left( u\right) }\right)
^{\dagger }\right) _{3i}\left( V_{uL}\widetilde{M}_{u}V_{uR}^{\dagger
}\right) _{i3}\left( V_{R}^{\left( u\right) }\right) _{3i} -\frac{\cos \al_{2}}{v_{\eta }} \left( \left( V_{L}^{\left( u\right) }\right) ^{\dagger }\right) _{3i}\left( V_{uL} \widetilde{M}_{u}V_{uR}^{\dagger }\right) _{i3}\left( V_{R}^{\left( u\right) }\right) _{3i}\right)^2 \,,
\eea
with $ q_1 = u, q_2 = c, i=1,2$. Hence, the branching ratios of $t \rightarrow q h_5 $ ($q=c, u$) are
\bea
Br(t\rightarrow  h_5 u) = \fr{\frac{g_{th_5u}^2}{4\pi}\fr{(m_t^2-m_{h_5}^2)^2}{2m_t m_{h_5}}}{\Ga_t}\,, \hs Br(t\rightarrow h_5 c ) = \fr{\frac{g_{th_5c}^2}{4\pi}\fr{(m_t^2-m_{h_5}^2)^2}{2m_t m_{h_5}}}{\Ga_t}\,.
\eea
Considering a benchmark scenario where the $h_5$ non SM scalar has a mass around $150$ GeV, we have numerically checked that the  branching ratios for the $t \rightarrow h_5 q$ decays (with $q=u,c$) can acquire values of the order of $10^{-3}$, which are within of the future experimental sensitivities.

In this section, we discuss the implications of the model in meson oscillations, in the $h \rightarrow \bar{b}b$, $h \rightarrow \bar{l}l$ decays as well as in the rare top decays $t \rightarrow c \ga$ and $t \rightarrow u \ga$. Furthermore, we also determine the couplings of the ALP $a$ and pseudoscalar $A_5$ and we provide the corresponding discussion.  

\subsection{SM like Higgs decays }

\subsubsection{SM like Higgs decays into two down-type quarks $h \rightarrow \bar{b}b$}
Using (\ref{decayratehff}), the decay rate of the process $h \rightarrow \bar{b}b$ is 
\bea
\Ga (h \rightarrow \bar{b}b) =  \int d\Ga = \fr{g_{hbb}^2}{8\pi}m_h\left(1-\fr{4m_b^2}{m_h^2}\right)^{\fr 3{2}}
\eea
with 
\bea
g_{hb\bar{b}}&=&\fr{\cos \alpha _{2}}{v_{\rho }}%
\left( \left( V_{L}^{\left( d\right) }\right)
^{\dagger }\right) _{33}\left( V_{dL}\widetilde{M}_{d}V_{dR}^{\dagger
}\right) _{33}\left( V_{R}^{\left( d\right) }\right) _{33} -\frac{\sin \al_{2}}{v_{\eta }} \left( \left( V_{L}^{\left( d\right) }\right) ^{\dagger }\right) _{33}\left( V_{dL} \widetilde{M}_{d}V_{dR}^{\dagger }\right) _{33}\left( V_{R}^{\left( d\right) }\right) _{33}\nn\\
&=&  \left(\fr{\cos \al_2  }{v_{\rho }}-\fr{\sin \al_2  }{v_{\eta }}\right)m_b=\left(\fr{\cos \al_2  }{\cos\al}-\fr{\sin \al_2  }{\sin\al}\right)\fr{m_b}{v}=\frac{\cos \al_2}{\sin \al}(\tan \al -\tan \al_2)\fr{m_b}{v}= \frac{2 \cos \al_2}{\cos \al}\frac{m_b}{v}\nn\\
&=&a_{h\bar{b}b}g_{hb\bar{b}}^{SM}.
\eea
where $a_{hb\bar{b}}$ is the deviation factor from the SM Higgs bottom quark coupling (in the SM this factor is unity). The experimental data constraint on the $a_{h\bar{b}b}$ parameter is given by:
\bea
a_{hb\bar{b}}^{exp}=0.91^{+0.17}_{-0.16} \,,
\eea

\subsubsection{SM like Higgs decays into two charged leptons $h \rightarrow \bar{l}l$}

Concerning the lepton sector, the Yukawa interaction for charged leptons are given by:
\be
-\mathcal{L}_{Y}^{\left( l\right) } = \sum_{a=1}^{3}\sum_{b=1}^{3}g_{ab}\overline{l}_{aL}\fr{v_{\rho }+R_{\rho }+iI_{\rho }}{%
	\sqrt{2}}l_{bR}+h.c \,. \label{LYl} 
\ee
Replacing Eq.(\ref{physstates}) and Eq.(\ref{physstateR}) in Eq. (\ref{LYl}), we get the Yukawa couplings of $h$ with leptons as below:
\bea
-\mathcal{L}_{Y}^{\left( l\right) } &\supset & \sum_{a=1}^{3}\sum_{b=1}^{3}\fr{g _{ab}\cos
	\alpha _{2}}{\sqrt{2}}\overline{l}_{aL}hl_{bR} \nn \\
& \supset &\sum_{a=1}^{3}\fr{\left( M_{l}\right) _{aa}\cos \al_{2}}{ v_{\rho }}\overline{l}_{aL}hl_{aR} \nn \\
& \supset &\sum_{a=1}^{3}\frac{v\cos \al_{2}}{v_{\rho }}\fr{\left(	M_{l}\right) _{aa}}{v}\overline{l}_{aL}hl_{aR}
. \label{LYhl}
\eea
\be
g_{h\bar{l}l}=\sum_{a=1}^{3}\fr{v\cos \alpha _{2}}{v_\rho}\fr{\left(M_{l}\right)_{aa}}{v}=a_{h\bar{l}l}g_{h\bar{l}l}^{SM}.
\ee
where $a_{h\bar{l}l}$ is the deviation of the $h\bar{l}l$ coupling with respect to the SM prediction (in the SM this factor is unity). 

\begin{figure}[h]
	\includegraphics[width=10cm, height=7cm]{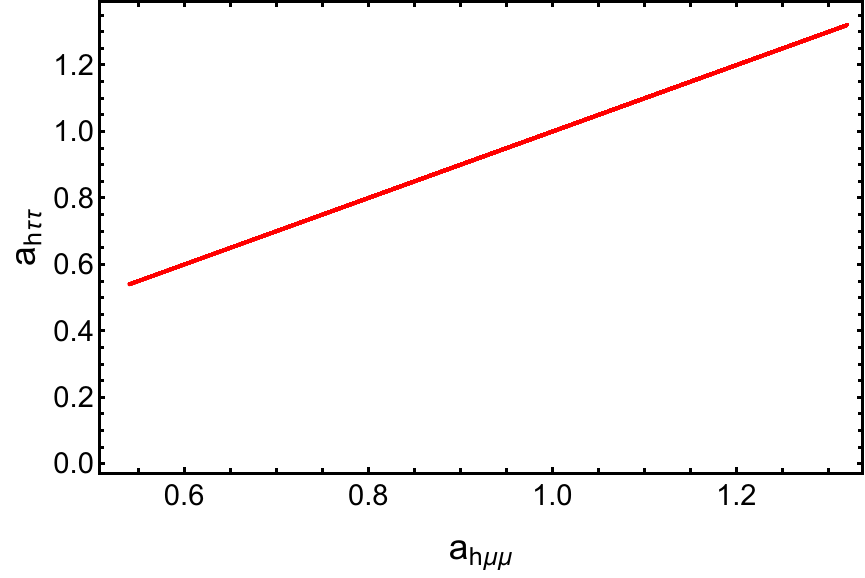}
	\caption{Correlation between the $a_{h\tau\tau}$ and $a_{h\mu\mu}$ parameters.}
	\label{ahcll}
\end{figure}

Using (\ref{decayratehff}), the decay rate of the process $h \rightarrow \mu \mu$ and $h \rightarrow \tau \tau$ are 
\bea
\Ga (h \rightarrow \mu \mu) =  \int d\Ga = \fr{g_{(h,\mu,\mu)}^2}{8\pi}m_h\left(1-\fr{4m_\mu^2}{m_h^2}\right)^{\fr 3{2}}= \left( \fr{v\cos \al_2}{v_\rho}\right)^2 \fr{m_\mu^2}{v^2} \fr{m_h}{8\pi}\left(1-\fr{4m_\mu^2}{m_h^2}\right)^{\fr 3{2}} = \left( \fr{\cos \al_2}{\cos \al}\right)^2 \fr{m_\mu^2}{v^2} \fr{m_h}{8\pi}\left(1-\fr{4m_\mu^2}{m_h^2}\right)^{\fr 3{2}}
\label{Gahmm}
\eea
\bea
\Ga (h \rightarrow \tau \tau) =  \int d\Ga = \fr{g_{(h,\tau,\tau)}^2}{8\pi}m_h\left(1-\fr{4m_\tau^2}{m_h^2}\right)^{\fr 3{2}}= \left( \fr{v\cos \al_2}{v_\rho}\right)^2 \fr{m_\tau^2}{v^2} \fr{m_h}{8\pi}\left(1-\fr{4m_\tau^2}{m_h^2}\right)^{\fr 3{2}} =\left( \fr{\cos \al_2}{\cos \al}\right)^2 \fr{m_\tau^2}{v^2} \fr{m_h}{8\pi}\left(1-\fr{4m_\tau^2}{m_h^2}\right)^{\fr 3{2}} 
\label{Gahtt}
\eea
From (\ref{Gahmm}) and (\ref{Gahtt}), one can get the constraints of the mixing angle $\al_2$ in this model. Using the following experimental allowed values of the parameters \cite{higgsll}:
\bea
a_{h\mu\mu}^{exp}=0.72^{+0.50}_{-0.72} \,,\hspace{2cm}
a_{h\tau\tau}^{exp}=0.93^{+0.13}_{-0.13}\,,
\eea
we can obtain plots where the allowed range of the mixing angle in the $CP$ even scalar sector is shown. Furthermore, we have found the our obtained values for the $a_{h \mu\mu,\tau\tau}$ parameters range from about $0.6$ up to about $1.2$, which is consistent with their current experimental bounds. This is shown in figure \ref{ahcll}, which displays a linear correlation between the $a_{h\tau\tau}$ and $a_{h\mu\mu}$ parameters.
\\
 
\begin{figure}[h]
	\includegraphics[width=10cm, height=7cm]{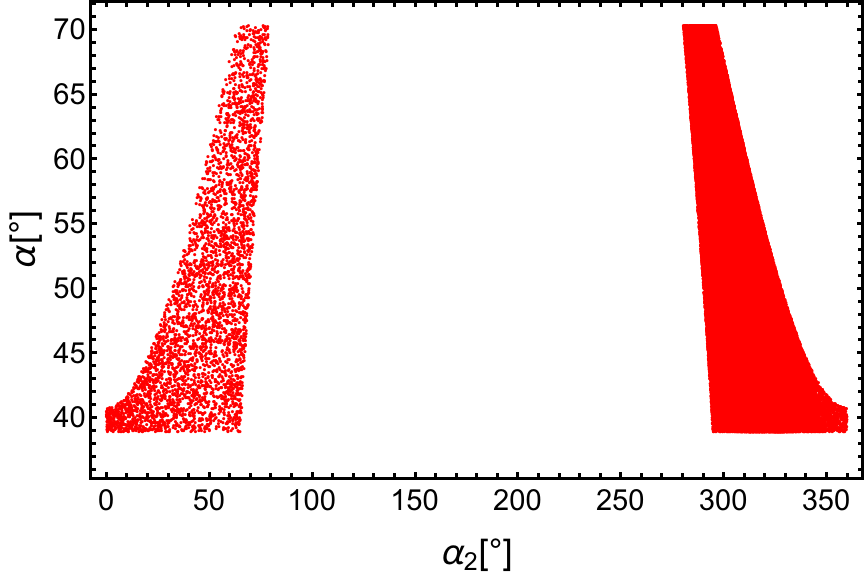}
	\caption{Correlation between the mixing angles $\al$ and $\al_2$ consistent with the experimental values of the $h \rightarrow \bar{\mu} \mu$, $h \rightarrow \bar{\tau} \tau$ and $h \rightarrow \bar{b} b$ decay rates.}
	\label{hcll}
\end{figure}

Requiring the consistency of the rates for the $h \rightarrow \bar{\mu} \mu$, $h \rightarrow \bar{\tau} \tau$ and $h \rightarrow \bar{b} b$ decays with their corresponding experimentally allowed ranges, we display in figure \ref{hcll} the correlation between the mixing angles $\al$ and $\al_2$. 

From the Fig.(\ref{hcll}), with $\alpha$ in range $38^o$  $\leq \al \leq 70^o$ , we get the following constraints for the mixing angle $\al_2$:
\be
0^o \,\leq \al_2 \leq 75^o \,,\hs \mbox{or}\,\hs 280^o \,  \leq \al_2 \leq 360^o \ \,.\label{conal2}
\ee

 We will use this constraint to analyze the meson oscillations of this model in the subsection below.

\subsection{Meson oscillations}
\label{FCNC}
 In this section, we analyze the consequences of the model under
consideration in the $K^{0}-\bar{K}^{0}$, $B_{d}^{0}-\bar{B}_{d}^{0}$ and $%
B_{s}^{0}-\bar{B}_{s}^{0}$ meson oscillations. These meson oscillations are
caused by flavor violating scalar and $Z^{\prime }$ interactions in the down
type quark sector. The  $K^{0}-\bar{K}^{0}$, $B_{d}^{0}-\bar{B}_{d}^{0}$ and
$B_{s}^{0}-\bar{B}_{s}^{0}$ meson mixings are described by the following
effective Hamiltonians:
\be
\mathcal{H}_{eff}^{\left( K^{0}-\bar{K}^{0}\right) }\mathcal{=}\fr{%
G_{F}^{2}m_{W}^{2}}{16\pi ^{2}}\sum_{i=1}^{3}C_{i}^{\left( K^{0}-\bar{K}%
^{0}\right) }\left( \mu \right) O_{i}^{\left( K^{0}-\bar{K}^{0}\right)
}\left( \mu \right) +\fr{4\sqrt{2}G_{F}c_{W}^{4}m_{Z}^{2}}{\left(
3-4s_{W}^{2}\right) m_{Z^{\prime }}^{2}}\left\vert \left( V_{DL}^{\ast
}\right) _{32}\left( V_{DL}\right) _{31}\right\vert ^{2}O_{4}^{\left( K^{0}-
\bar{K}^{0}\right) },
\label{K}
\ee
\be
\mathcal{H}_{eff}^{\left( B_{d}^{0}-\bar{B}_{d}^{0}\right) }\mathcal{=}\fr{%
G_{F}^{2}m_{W}^{2}}{16\pi ^{2}}\sum_{i=1}^{3}C_{i}^{\left( B_{d}^{0}-\bar{B}%
_{d}^{0}\right) }\left( \mu \right) O_{i}^{\left( B_{d}^{0}-\bar{B}%
_{d}^{0}\right) }\left( \mu \right) +\fr{4\sqrt{2}G_{F}c_{W}^{4}m_{Z}^{2}}{%
\left( 3-4s_{W}^{2}\right) m_{Z^{\prime }}^{2}}\left\vert \left(
V_{DL}^{\ast }\right) _{31}\left( V_{DL}\right) _{33}\right\vert
^{2}O_{4}^{\left( B_{d}^{0}-\bar{B}_{d}^{0}\right) },
\label{Bd}
\ee
\be
\mathcal{H}_{eff}^{\left( B_{s}^{0}-\bar{B}_{s}^{0}\right) }\mathcal{=}\fr{%
G_{F}^{2}m_{W}^{2}}{16\pi ^{2}}\sum_{i=1}^{3}C_{i}^{\left( B_{s}^{0}-\bar{B}%
_{s}^{0}\right) }\left( \mu \right) O_{i}^{\left( B_{s}^{0}-\bar{B}%
_{s}^{0}\right) }\left( \mu \right) +\fr{4\sqrt{2}G_{F}c_{W}^{4}m_{Z}^{2}}{%
\left( 3-4s_{W}^{2}\right) m_{Z^{\prime }}^{2}}\left\vert \left(
V_{DL}^{\ast }\right) _{32}\left( V_{DL}\right) _{33}\right\vert
^{2}O_{4}^{\left( B_{s}^{0}-\bar{B}_{s}^{0}\right) },
\label{Bs}
\ee
where $V_{DL}$ is the rotation matrix that diagonalizes $M_DM_D^{\dagger}$ according to $V_{DL}^{\dagger}M_DM_D^{\dagger}V_{DL}=diag(m^2_d,m^2_s,m^2_b)$ being $M_D$ the SM down type quark mass matrix. Furthermore, the operators appearing in Eqs. (\ref{K}), (\ref{Bd}) and (\ref{Bs}) are given by:
\bea 
O_{1}^{\left( K^{0}-\bar{K}^{0}\right) } &=&\left( \overline{s}P_{L}d\right)
\left( \overline{s}P_{L}d\right) ,\hspace{0.7cm}\hspace{0.7cm}O_{2}^{\left(
	K^{0}-\bar{K}^{0}\right) }=\left( \overline{s}P_{R}d\right) \left( \overline{%
	s}P_{R}d\right) ,\hspace{0.7cm}  \label{op3f} \\
O_{3}^{\left( K^{0}-\bar{K}^{0}\right) } &=&\left( \overline{s}P_{L}d\right)
\left( \overline{s}P_{R}d\right) ,\hspace{0.7cm}\hspace{0.7cm}O_{4}^{\left(
	K^{0}-\bar{K}^{0}\right) }=\left( \overline{s}\gamma _{\mu }P_{L}d\right)
\left( \overline{s}\gamma ^{\mu }P_{L}d\right) , \\
O_{1}^{\left( B_{d}^{0}-\bar{B}_{d}^{0}\right) } &=&\left( \overline{d}%
P_{L}b\right) \left( \overline{d}P_{L}b\right) ,\hspace{0.7cm}\hspace{0.7cm}%
O_{2}^{\left( B_{d}^{0}-\bar{B}_{d}^{0}\right) }=\left( \overline{d}%
P_{R}b\right) \left( \overline{d}P_{R}b\right) ,\hspace{0.7cm} \\
O_{3}^{\left( B_{d}^{0}-\bar{B}_{d}^{0}\right) } &=&\left( \overline{d}%
P_{L}b\right) \left( \overline{d}P_{R}b\right) ,\hspace{0.7cm}\hspace{0.7cm}%
O_{4}^{\left( B_{d}^{0}-\bar{B}_{d}^{0}\right) }=\left( \overline{d}\gamma
_{\mu }P_{L}b\right) \left( \overline{d}\gamma ^{\mu }P_{L}b\right) , \\
O_{1}^{\left( B_{s}^{0}-\bar{B}_{s}^{0}\right) } &=&\left( \overline{s}%
P_{L}b\right) \left( \overline{s}P_{L}b\right) ,\hspace{0.7cm}\hspace{0.7cm}%
O_{2}^{\left( B_{s}^{0}-\bar{B}_{s}^{0}\right) }=\left( \overline{s}%
P_{R}b\right) \left( \overline{s}P_{R}b\right) , \\
O_{3}^{\left( B_{s}^{0}-\bar{B}_{s}^{0}\right) } &=&\left( \overline{s}%
P_{L}b\right) \left( \overline{s}P_{L}b\right) ,\hspace{0.7cm}\hspace{0.7cm}%
O_{4}^{\left( B_{s}^{0}-\bar{B}_{s}^{0}\right) }=\left( \overline{s}\gamma
_{\mu }P_{L}b\right) \left( \overline{s}\gamma ^{\mu }P_{L}b\right) ,
\eea 
and the Wilson coefficients read:
\bea
C_{1}^{\left( K^{0}-\bar{K}^{0}\right) } &=&\fr{16\pi ^{2}}{%
G_{F}^{2}m_{W}^{2}}\left( \fr{g_{h\overline{s}_{R}d_{L}}^{2}}{m_{h}^{2}}+%
\fr{g_{h_5\overline{s}_{R}d_{L}}^{2}}{m_{h_5}^{2}}-\fr{g_{A_5%
\overline{s}_{R}d_{L}}^{2}}{m_{A_5}^{2}}\right) , \\
C_{2}^{\left( K^{0}-\bar{K}^{0}\right) } &=&\fr{16\pi ^{2}}{%
G_{F}^{2}m_{W}^{2}}\left( \fr{g_{h\overline{s}_{L}d_{R}}^{2}}{m_{h}^{2}}+%
\fr{g_{h_5\overline{s}_{L}d_{R}}^{2}}{m_{h_5}^{2}}-\fr{g_{A_5%
\overline{s}_{L}d_{R}}^{2}}{m_{A_5}^{2}}\right) , \\
C_{3}^{\left( K^{0}-\bar{K}^{0}\right) } &=&\fr{16\pi ^{2}}{%
G_{F}^{2}m_{W}^{2}}\left( \fr{g_{h\overline{s}_{R}d_{L}}g_{h\overline{s}%
_{L}d_{R}}}{m_{h}^{2}}+\fr{g_{h_5\overline{s}_{R}d_{L}}g_{h_5\overline{%
s}_{L}d_{R}}}{m_{h_5}^{2}}-\fr{g_{A_5\overline{s}_{R}d_{L}}g_{A_5%
\overline{s}_{L}d_{R}}}{m_{A_5}^{2}}\right) ,
\eea %
\bea
C_{1}^{\left( B_{d}^{0}-\bar{B}_{d}^{0}\right) } &=&\fr{16\pi ^{2}}{%
G_{F}^{2}m_{W}^{2}}\left( \fr{g_{h\overline{d}_{R}b_{L}}^{2}}{m_{h}^{2}}+%
\fr{g_{h_5\overline{d}_{R}b_{L}}^{2}}{m_{h_5}^{2}}-\fr{g_{A_5%
\overline{d}_{R}b_{L}}^{2}}{m_{A_5}^{2}}\right) , \\
C_{2}^{\left( B_{d}^{0}-\bar{B}_{d}^{0}\right) } &=&\fr{16\pi ^{2}}{%
G_{F}^{2}m_{W}^{2}}\left( \fr{g_{h\overline{d}_{L}b_{R}}^{2}}{m_{h}^{2}} +%
\fr{g_{h_5\overline{d}_{L}b_{R}}^{2}}{m_{h_5}^{2}} -\fr{g_{A_5%
\overline{d}_{L}b_{R}}^{2}}{m_{A_5}^{2}}\right) , \\
C_{3}^{\left( B_{d}^{0}-\bar{B}_{d}^{0}\right) } &=&\fr{16\pi ^{2}}{%
G_{F}^{2}m_{W}^{2}}\left( \fr{g_{h\overline{d}_{R}b_{L}}g_{h\overline{d}%
_{L}b_{R}}}{m_{h}^{2}}+\fr{g_{h_5\overline{d}_{R}b_{L}}g_{h_5\overline{%
d}_{L}b_{R}}}{m_{h_5}^{2}}-\fr{g_{A_5\overline{d}_{R}b_{L}}g_{A_5%
\overline{d}_{L}b_{R}}}{m_{A_5}^{2}}\right) ,
\eea %
\bea
C_{1}^{\left( B_{s}^{0}-\bar{B}_{s}^{0}\right) } &=&\fr{16\pi ^{2}}{%
G_{F}^{2}m_{W}^{2}}\left( \fr{g_{h\overline{s}_{R}b_{L}}^{2}}{m_{h}^{2}}+%
\fr{g_{h_5\overline{s}_{R}b_{L}}^{2}}{m_{h_5}^{2}}-\fr{g_{A_5%
\overline{s}_{R}b_{L}}^{2}}{m_{A_5}^{2}}\right) , \\
C_{2}^{\left( B_{s}^{0}-\bar{B}_{s}^{0}\right) } &=&\fr{16\pi ^{2}}{%
G_{F}^{2}m_{W}^{2}}\left( \fr{g_{h\overline{s}_{L}b_{R}}^{2}}{m_{h}^{2}}+%
\fr{g_{h_5\overline{s}_{L}b_{R}}^{2}}{m_{h_5}^{2}}-\fr{g_{A_5%
\overline{s}_{L}b_{R}}^{2}}{m_{A_5}^{2}}\right) , \\
C_{3}^{\left( B_{s}^{0}-\bar{B}_{s}^{0}\right) } &=&\fr{16\pi ^{2}}{%
G_{F}^{2}m_{W}^{2}}\left( \fr{g_{h\overline{s}_{R}b_{L}}g_{h\overline{s}%
_{L}b_{R}}}{m_{h}^{2}}+\fr{g_{h_5\overline{s}_{R}b_{L}}g_{h_5\overline{%
s}_{L}b_{R}}}{m_{h_5}^{2}}-\fr{g_{A_5\overline{s}_{R}b_{L}}g_{A_5%
\overline{s}_{L}b_{R}}}{m_{A_5}^{2}}\right) ,
\eea
with $g_{abc}$  are the couplings between the scalar $a=h, h_5, A_5$ and down-type quarks $b=\bar{d}_{L,R}^i$, $c=d_{L,R}^j$, $i,j=1,2,3$, $i \neq j$.\\ 
On the other hand, the $K-\bar{K}$, $B_{d}^{0}-\bar{B}_{d}^{0}$ and $%
B_{s}^{0}-\bar{B}_{s}^{0}$\ mass splittings are given by:
\be
\De m_{K}=\left( \De m_{K}\right) _{SM}+\De m_{K}^{\left( NP\right)
},\hs\De m_{B_{d}}=\left( \De m_{B_{d}}\right) _{SM}+\De
m_{B_{d}}^{\left( NP\right) },\hs\Delta m_{B_{s}}=\left( \De
m_{B_{s}}\right) _{SM}+\De m_{B_{s}}^{\left( NP\right) },  \label{Deltam}
\ee %
where $\left( \De m_{K}\right) _{SM}$, $\left( \De m_{B_{d}}\right)
_{SM}$ and $\left( \De m_{B_{s}}\right) _{SM}$ are the SM contributions,
whereas $\De m_{K}^{\left( NP\right) }$ , $\De m_{B_{d}}^{\left(
NP\right) }$ and $\left( \De m_{B_{s}}\right) _{SM}$ are new physics
contributions.
\newline
In the model under consideration, the new physics contributions to the meson differences are
given by:
\bea
\De m_{K}^{\left( NP\right) } &=&\fr{4\sqrt{2}G_{F}c_{W}^{4}m_{Z}^{2}}{%
\left( 3-4s_{W}^{2}\right) m_{Z^{\prime }}^{2}}\left\vert \left(
V_{DL}^{\ast }\right) _{32}\left( V_{DL}\right) _{31}\right\vert
^{2}f_{K}^{2}B_{K}\eta _{K}m_{K}  \crn
&&+\fr{G_{F}^{2}m_{W}^{2}}{6\pi ^{2}}m_{K}f_{K}^{2}\eta _{K}B_{K}\left[
P_{2}^{\left( K^{0}-\bar{K}^{0}\right) }C_{3}^{\left( K^{0}-\bar{K}%
^{0}\right) }+P_{1}^{\left( K^{0}-\bar{K}^{0}\right) }\left( C_{1}^{\left(
K^{0}-\bar{K}^{0}\right) }+C_{2}^{\left( K^{0}-\bar{K}^{0}\right) }\right) %
\right] \,,  \nn
\eea%
\bea
\De m_{B_{d}}^{\left( NP\right) } &=&\fr{4\sqrt{2}%
G_{F}c_{W}^{4}m_{Z}^{2}}{\left( 3-4s_{W}^{2}\right) m_{Z^{\prime }}^{2}}%
\left\vert \left( V_{DL}^{\ast }\right) _{31}\left( V_{DL}\right)
_{33}\right\vert ^{2}f_{B_{d}}^{2}B_{B_{d}}\eta _{B_{d}}m_{B_{d}}  \crn
&&+\fr{G_{F}^{2}m_{W}^{2}}{6\pi ^{2}}m_{B_{d}}f_{B_{d}}^{2}\eta
_{B_{d}}B_{B_{d}}\left[ P_{2}^{\left( B_{d}^{0}-\bar{B}_{d}^{0}\right)
}C_{3}^{\left( B_{d}^{0}-\bar{B}_{d}^{0}\right) }+P_{1}^{\left( B_{d}^{0}-%
\bar{B}_{d}^{0}\right) }\left( C_{1}^{\left( B_{d}^{0}-\bar{B}%
_{d}^{0}\right) }+C_{2}^{\left( B_{d}^{0}-\bar{B}_{d}^{0}\right) }\right) %
\right] \,,  \nn
\eea%
\bea
\De m_{B_{s}}^{\left( NP\right) } &=&\fr{4\sqrt{2}%
G_{F}c_{W}^{4}m_{Z}^{2}}{\left( 3-4s_{W}^{2}\right) m_{Z^{\prime }}^{2}}%
\left\vert \left( V_{DL}^{\ast }\right) _{32}\left( V_{DL}\right)
_{33}\right\vert ^{2}f_{B_{s}}^{2}B_{B_{s}}\eta _{B_{s}}m_{B_{s}}  \crn
&&+\fr{G_{F}^{2}m_{W}^{2}}{6\pi ^{2}}m_{B_{s}}f_{B_{s}}^{2}\eta
_{B_{s}}B_{B_{s}}\left[ P_{2}^{\left( B_{s}^{0}-\bar{B}_{s}^{0}\right)
}C_{3}^{\left( B_{s}^{0}-\bar{B}_{s}^{0}\right) }+P_{1}^{\left( B_{s}^{0}-%
\bar{B}_{s}^{0}\right) }\left( C_{1}^{\left( B_{s}^{0}-\bar{B}%
_{s}^{0}\right) }+C_{2}^{\left( B_{s}^{0}-\bar{B}_{s}^{0}\right) }\right) %
\right]   \nn
\eea%
Using the following parameters \cite{Zyla:2020zbs}:
\bea
\left(\De m_{K}\right)_{\exp }&=&\left( 3.484\pm 0.006\right) \times 10^{-12}\, \rm{MeV},\hspace{1.5cm%
}\left( \De m_{K}\right) _{SM}=3.483\times 10^{-12}\, \rm{MeV}  \crn
f_{K} &=&155.7\, \rm{MeV},\hspace{1.5cm}B_{K}=0.85,\hspace{1.5cm}\eta _{K}=0.57,
\crn
P_{1}^{\left( K^{0}-\bar{K}^{0}\right) } &=&-9.3,\hspace{1.5cm}P_{2}^{\left(
K^{0}-\bar{K}^{0}\right) }=30.6,\hspace{1.5cm}m_{K}=\left(497.611\pm 0.013\right)\, \rm{MeV},\hspace{1.5cm}
\eea %
\bea
\left( \De m_{B_{d}}\right) _{\exp } &=&\left(3.334\pm 0.013\right)
\times 10^{-10}\, \rm{MeV},\hspace{1.5cm}\left( \De m_{B_{d}}\right)
_{SM}=\left(3.653\pm 0.037\pm 0.019\right)\times 10^{-10}\, \rm{MeV},  \crn
f_{B_{d}} &=&188\, \rm{MeV},\hspace{1.5cm}B_{B_{d}}=1.26,\hspace{1.5cm}\eta
_{B_{d}}=0.55,  \crn
P_{1}^{\left( B_{d}^{0}-\bar{B}_{d}^{0}\right) } &=&-0.52,\hspace{1.5cm}%
P_{2}^{\left( B_{d}^{0}-\bar{B}_{d}^{0}\right) }=0.88,\hspace{1.5cm}%
m_{B_{d}}=\left(5279.65\pm 0.12\right)\,\rm{MeV},\hspace{1.5cm}
\eea %
\bea
\left( \De m_{B_{s}}\right) _{\exp } &=&\left(1.1683\pm 0.0013\right) \times 10^{-8}\, \rm{MeV},\hspace{1.5cm}\left( \De m_{B_{s}}\right)
_{SM}=\left(1.1577\pm 0.022\pm 0.051\right) \times 10^{-8}\, \rm{MeV},  \crn
f_{B_{s}} &=&225\, \rm{MeV},\hspace{1.5cm}B_{B_{s}}=1.33,\hspace{1.5cm}\eta
_{B_{s}}=0.55,  \crn
P_{1}^{\left( B_{s}^{0}-\bar{B}_{s}^{0}\right) } &=&-0.52,\hspace{1.5cm}%
P_{2}^{\left( B_{s}^{0}-\bar{B}_{s}^{0}\right) }=0.88,\hspace{1.5cm}%
m_{B_{s}}=\left(5366.9\pm 0.12\right)\, \rm{MeV},\hspace{1.5cm}
\eea%

\begin{figure}[h]
	\includegraphics[width=10cm, height=8cm]{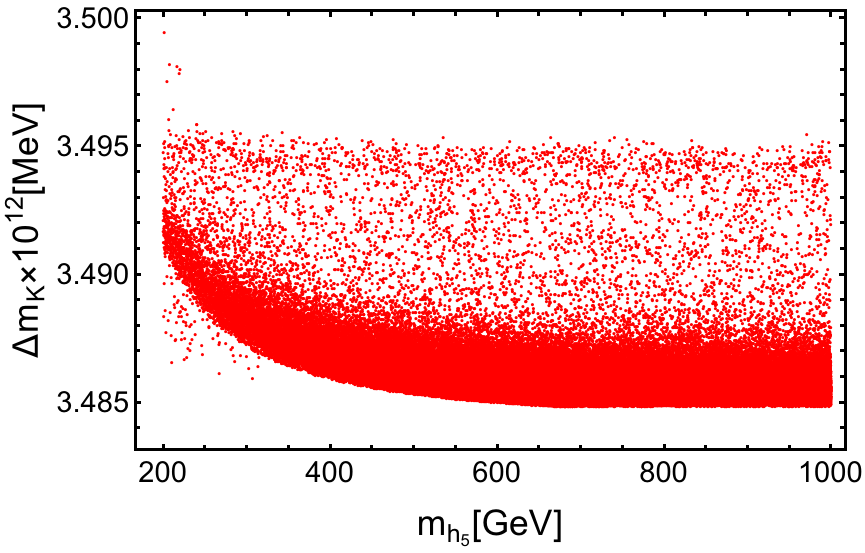}
	\caption{Correlation of the $\De m_{K}$ meson mass splitting with the heayy $CP$ even scalar mass $m_{h^5}$. }
	\label{mesons}
\end{figure}

\begin{figure}[h]
	\includegraphics[width=10cm, height=8cm]{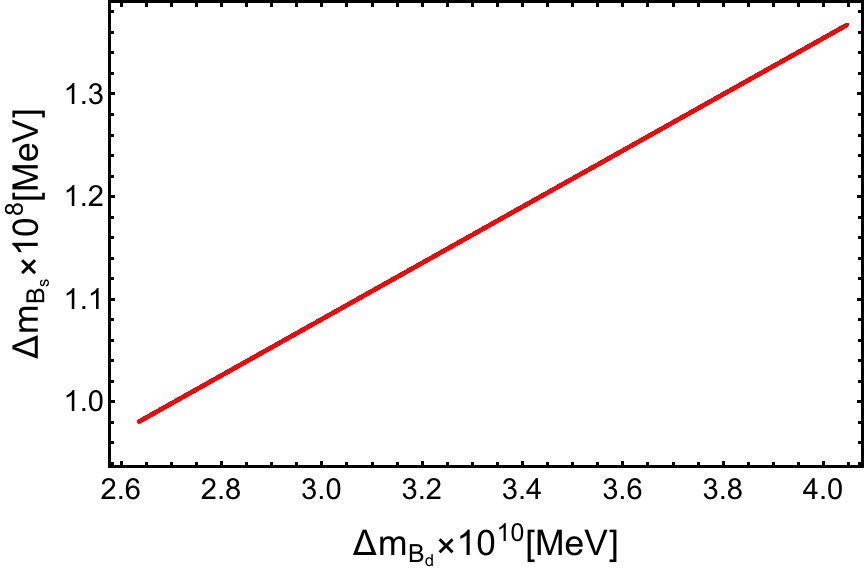}
	\caption{Correlation between $\De m_{B_{d}}$ and $\De m_{B_{s}}$ meson mass splittings. }
	\label{Bmesons}
\end{figure}

\begin{figure}[h]
	\includegraphics[width=10cm, height=7.5cm]{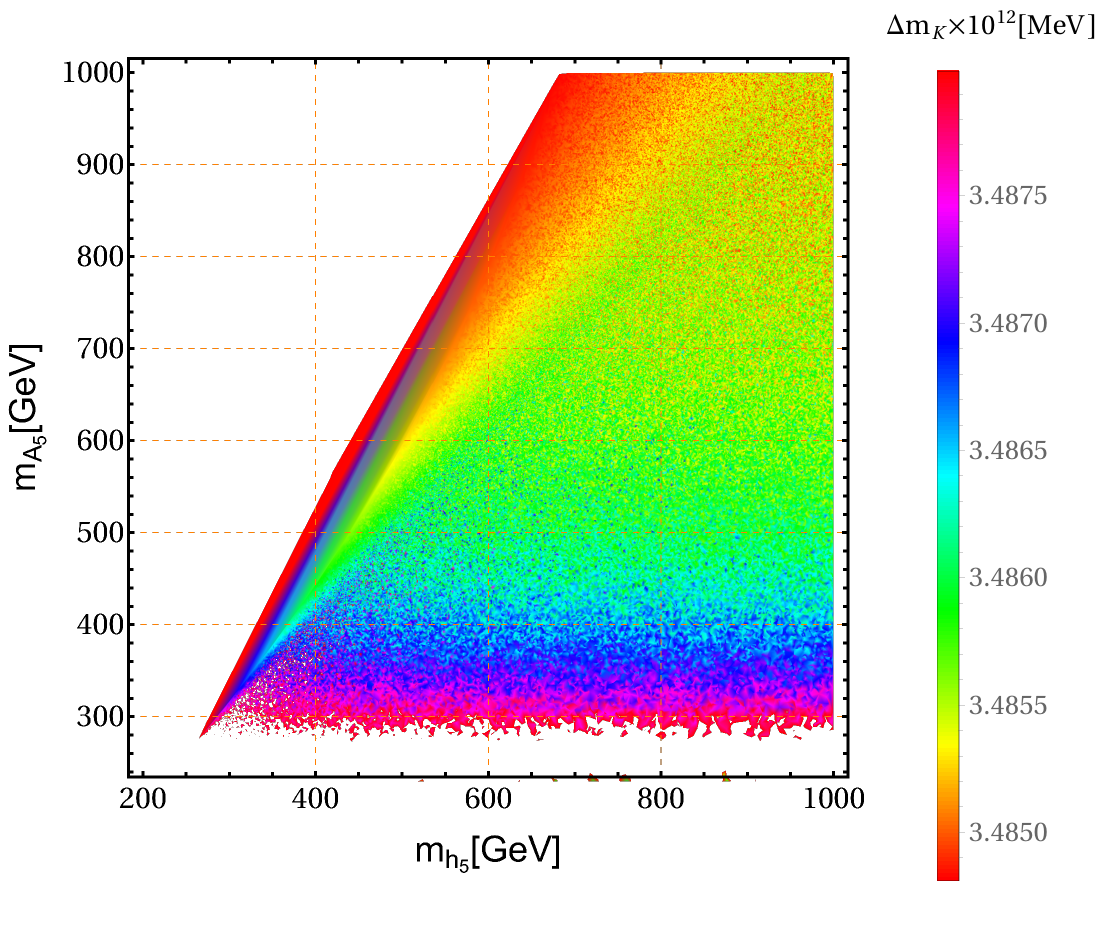}
	\caption{Allowed region in the $m_{A_5}-m_{h_5}$ plane consistent with the constraints on $\Delta m_K$, $\Delta m_{B_d}$ and $\Delta m_{B_s}$ meson mass splittings.}
	\label{densitymA5mh5deltamK}
\end{figure}

We plot in Fig. \ref{mesons} the correlation between of the $\De m_{K}$ meson mass splitting with the non SM $CP$ even scalar mass $m_{h_5}$, whereas in Fig. (\ref{densitymA5mh5deltamK}) we display the allowed region in the $m_{A_5}-m_{h_5}$ plane consistent with the constraints on $\De m_K$, $\Delta m_{B_d}$ and $\Delta m_{B_s}$ meson mass splittings, whose obtained values are within the experimentally allowed range. As seen from Figs. (\ref{mesons}) and (\ref{densitymA5mh5deltamK}), if one keeps the other parameters fixed, an increase of the non SM $CP$ even scalar mass $m_{h_5}$ yields a decrease of the $\De m_{K}$ meson mass difference. Besides that, Fig. (\ref{mesons}) indicates that the number of solutions consistent with the meson oscillation constraints is increased when the mass $m_{h_5}$ of the non SM $CP$ even scalar $h_5$ acquires larger values close to the TeV scale. This is due to the fact that the scalar contributions to the meson mass splittings are inversely proportional to the square of the scalar and pseudoscalar masses $m_{h_5}$ and $m_{A_5}$, then making easier to find more solutions consistent with the meson oscillation constraints in the large mass region than in the low mass region of the non SM scalars.
 Here the $CP$ even and $CP$ odd scalar masses have been varied in the ranges 
$200 \, \textrm{GeV} \leq m_{h_5} \leq  1$  TeV and $100 \, \textrm{GeV} \leq m_{A_5}\leq 1$ TeV, respectively. In our numerical analysis we have varied the mixing angles $\alpha$, $\alpha_2$ in a range of values consistent with the experimental constraints of the $h\tau\bar{\tau}$, $h\mu\bar{\mu}$ and and $hb\bar{b}$ couplings (being $h$ is the $126$ SM like Higgs boson) as well as with the meson oscillation constraints. Besides that, the VEV $v_{\eta}$ of the neutral component of the $SU(3)_L$ scalar triplet have been varied in window around $200$ GeV, which is consistent with the experimental constraints on meson mass splittings. Moreover, 
%
  we have considered a simplified benchmark scenario of real down type quark sector parameters so that the $CP$ violation in the quark sector entirely arises from the up type quark sector. Furthermore, we have set the $Z^{\prime }$ mass to be equal to $6$~TeV, which is consistent with the constraints arising from collider searches \cite{ATLAS:2019erb,CMS:2021ctt}. Moreover, a linear correlation between $\De m_{B_{d}}$ and $\De m_{B_{s}}$ meson mass splittings is displayed in figure \ref{Bmesons}.
  As seen from Figs \ref{mesons} and \ref{densitymA5mh5deltamK}, the model under consideration successfully fulfills the constraints arising from the meson oscillation experimental data and the obtained meson mass differences $K-\bar{K}$, $B_{d}^{0}-\bar{B}_{d}^{0}$ and $B_{s}^{0}-\bar{B}_{s}^{0}$ reach values within the reach of experimental sensitivity. Given that we are considering the case of real down type quark sector parameters, the constraints that are usually imposed on any possible new contributions to the $K^{0}-\bar{K}^{0}$, $B_{d}^{0}-\bar{B}_{d}^{0}$ and $B_{s}^{0}-\bar{B}_{s}^{0}$ meson oscillations arising from $CP$-violating processes are not relevant for our case.

\subsection{Rare top decays $t \rightarrow c \ga$ and $t \rightarrow u \ga$ with flavor changing neutral scalar interactions}

In this section we discuss about the implications of the model under consideration in the rare top quark decays $t\rightarrow u\ga$ and $t\rightarrow c\ga$. In the SM these decays have very tiny branching ratios, however in extensions of the SM, like the 331 model considered in this paper, the branching ratios of these decays can be significantly enhanced with respect to the SM prediction. This is due to the flavor changing neutral scalar interactions in the quark sector, which provide the dominant contributions to the are top quark decays $t\rightarrow u\ga$ and $t\rightarrow c\ga$.

The one loop Feynman diagram is with a neutral Higgs boson in the internal line. This diagram shows the flavor changing neutral scalar contribution \cite{tcg}. The rare top quark decays $t\rightarrow u\ga$ and $t\rightarrow c\ga$ also receive contributions from electrically charged scalars and down type quarks, however those contributions are subleading. Thus, the decay rate for the $t\rightarrow c\ga$ and $t\rightarrow u\ga$ processes have the form \cite{tcg}:  
\bea 
\Ga \left( t\rightarrow c\ga \right) &=&\fr{\al
	G_{F}m_{t}^{3}\left\vert y_{hct}\right\vert ^{2}}{192\pi ^{4}}\left\vert
\left( f_{1}\left( \fr{m_{h}}{m_{t}}\right) +f_{2}\left( \fr{m_{h}}{m_{t}%
}\right) \right) A_{h}B_{h}+\left( f_{1}\left( \fr{m_{h_{5}}}{m_{t}}%
\right) +f_{2}\left( \fr{m_{h_{5}}}{m_{t}}\right) \right)
A_{h_{5}}B_{h_{5}}\right\vert ^{2}  \crn 
\Ga \left( t\rightarrow u\ga\right) &=&\fr{\al
	G_{F}m_{t}^{3}\left\vert y_{hut}\right\vert ^{2}}{192\pi ^{4}}\left\vert
\left( f_{1}\left( \fr{m_{h}}{m_{t}}\right) +f_{2}\left( \fr{m_{h}}{m_{t}%
}\right) \right) A_{h}B_{h}+\left( f_{1}\left( \fr{m_{h_{5}}}{m_{t}}%
\right) +f_{2}\left( \fr{m_{h_{5}}}{m_{t}}\right) \right)
A_{h_{5}}B_{h_{5}}\right\vert ^{2}
\eea 
where:
\bea 
A_{h} &=&-\fr{\sin \alpha _{2}}{\sin \beta },\hspace{1cm}A_{h_{5}}=\fr{%
	\cos \al_{2}}{\sin \beta },  \crn
B_{h} &=&\fr{\sin \al_{2}}{\sin \beta }+\frac{\cos \al_{2}}{\cos
	\beta },\hspace{1cm}B_{h_{5}}=-\fr{\cos \al_{2}}{\sin \beta }+\fr{%
	\sin \al_{2}}{\cos \beta }.
\eea 
and the loop integrals are given by:
\bea 
f_{1}\left( z\right)  &=&\int_{0}^{1}dx\int_{0}^{1-x}dy\fr{x\left(
	x+y-1\right) }{x^{2}+xy-\left( 2-z^{2}\right) +1},  \crn
f_{2}\left( z\right)  &=&\int_{0}^{1}dx\int_{0}^{1-x}dy\fr{x-1}{%
	x^{2}+xy-\left( 2-z^{2}\right) +1}
\eea 
It is worth mentioning that, in order to simplify our analysis, we have considered a simplified benchmark scenario where the neutral $CP$ odd scalar $A_5$ has a mass close to the TeV scale, whereas the $CP$ even neutral scalar $h_5$ has a mass in the range $100$ GeV$\leq m_{h_5}\leq$$200$ GeV. Then, in this scenario, the leading contributions to the $t\rightarrow u\ga$ and $t\rightarrow c\ga$ decays will arise from the virtual exchange of the top quark and neutral $CP$ even scalars $h$ and $h_5$, being $h$ the $126$ GeV SM like Higgs boson. Furthermore, we have varied the flavor changing top quark Yukawa couplings $y_{hct}$ and $y_{hut}$ in the range $10^{-2}$ GeV$\leq y_{hct},y_{hut}\leq$$1.2\times 10^{-2}$. The branching ratio for the rare top quark decays $t\rightarrow c\ga$ and $t\rightarrow u\ga$ are given by:
\bea 
Br\left( t\rightarrow c\gamma \right)=\fr{\Ga \left( t\rightarrow c\ga \right)}{\Ga_{top}},\hspace{2cm}Br\left( t\rightarrow u\ga  \right)=\frac{\Ga \left( t\rightarrow u\ga \right)}{\Ga_{top}},
\eea 
where $\Ga_{top}=1.42^{+0.19}_{-0.15}$ GeV is the total top quark decay width. We have numerically checked that the branching ratios for the $t\rightarrow c\ga$ and $t\rightarrow u\ga$ decays acquire values of the order of $10^{-10}$, several orders of magnitude lower than their corresponding experimental upper bounds of $2.2\times 10^{-4}$ and $6.1\times 10^{-5}$, respectively. On the other hand, our obtained values for the $t\rightarrow c\ga$ and $t\rightarrow u\ga$ decay branching ratios are $4$ and $6$ orders of magnitude larger than their corresponding SM values of $4.6\times 10^{-14}$ and $3.7\times 10^{-16}$, respectively.

	\subsection{Couplings of ALP $a$ and pseudoscalar $A_5$}
	\subsubsection{Coupings with exotic quarks}

Due to $Z_2$ symmetry, all terms containing the Yukawa interactions of ordinary quarks with ALP $a$  are forbidden.  The ALP $a$ just interact with exotic quarks. Hence, one has
	\bea
	\mathcal{L}_a^Y &=&
\sqrt{2}	i a\, \sin \theta_\phi \sin \theta_3 \left(
	\fr{m_U }{v_\chi} \bar{U} \ga_5 U -  \sum\limits_{\al =1}^{2}\fr{m_{D_\al} }{v_\chi}\bar{D}_{\al } \ga_5 D_{\al } \right)  \,	\label{axcouplingf}\,. \eea
About the interactions between the pseudoscalar $A_5$ with quarks in the model, this $A_5$ interacts with not only exotic quarks but also ordinary quarks. The Yukawa interaction between $A_5$ with exotic quarks can be defined by the equation below:
	\bea
	\mathcal{L}_{A_5}^Y  &\approx&
\sqrt{2}	i A_5\, \cos \theta_\phi \sin \theta_3 \left(
	-\fr{m_U }{v_\chi} \bar{U} \ga_5 U + \sum\limits_{\al =1}^{2}\fr{m_{D_\al} }{v_\chi}\bar{D}_{\al } \ga_5 D_{\al } \right)\,. \label{A5couplingsf}
	\eea
So, the ALP interacts only with exotic quarks with tiny strength ($\propto \sin \theta_\phi \sin \theta_3 $). This property is suitable with one of properties of dark matter. This is the reason why ALP $a$ can be regarded as a candidate of dark matter.  Remember that $\sin \theta_3$ is also very small, so the strength of interactions between the pseudoscalar $A_5$ and exotic quarks are also tiny($\propto  \sin \theta_3 $). From Eq.(\ref{axcouplingf}) and Eq.(\ref{A5couplingsf}), one gets the couplings of ALP $a$ and pseudoscalar $A_5$ with exotic quarks as below:
	\bea
	g_{a}^{Q_i}= i\ga_5\sqrt{2} \sin \theta_\phi \sin \theta_3 \frac{m_{q_i}}{v_\chi}\,, \label{gaQ}\\
	g_{A_5}^{Q_i}= i\ga_5\sqrt{2} \cos \theta_\phi \sin \theta_3 \frac{m_{q_i}}{v_\chi}\,, \label{gA5Q}
	\eea
with $i=\al,3$, $\al=1,2$, $Q_\al = D_\al$, $Q_3 = U$. From Eq.(\ref{gaQ}) and Eq.(\ref{gA5Q}), we have:
	\be g_{a}^{Q_i} \ll g_{A_5}^{Q_i}
		\ee

	\subsubsection{Coupings with SM like Higgs $h$ and new light Higgs $h_5$}
The coupling of $h$ and two ALP $a$ is defined from (\ref{Vhaa2}) as below:
\be
g_{haa} \approx \fr{v_\rho v_\eta}{2\sqrt2} \left(
\frac{\la_6 \la_{12}}{\sqrt{V_{236}^2+(\la_3 v_\rho^2 - \la_2 v_\eta^2)V_{236}}}-\la_{13}\sqrt{V_{236}+\la_3 v_\rho^2 - \la_2 v_\eta^2}
\right)\,, \label{haa}
\ee
with $V_{236}=\sqrt{\left(\la_2 v_\eta^2-\la_3 v_\rho^2\right)^2+\la_6^2 v_\eta^2 v_\rho^2}$.\\
We also get the coupling of $h$ and two pseudoscalar $A_5$ from (\ref{VhA5A5}):
\be
g_{hA_5 A_5} \approx  \fr{1}{2\sqrt2} \left(v_\rho(2\la_3 v_\eta^2 +\la_6 v_\rho^2)\sqrt{\frac{V_{236}-\la_3 v_\rho^2 +\la_2 v_\eta^2}{V_{236}}} -v_\eta(2\la_2 v_\rho^2 +\la_6 v_\eta^2)\sqrt{\fr{V_{236}+\la_3 v_\rho^2 -\la_2 v_\eta^2}{V_{236}}}
\right)\,. \label{hA5A5}
\ee
Similarly with the new light Higgs $h_5$, use (\ref{Vh5aa}) and (\ref{Vh5A5A5}) one gets:
\be
g_{h_5aa} \approx  \fr{1}{2\sqrt2}v_\rho \left(\la_{12}\sqrt{V_{236}+\la_3^2 v_\rho^2 -\la_2 v_\eta^2}+\fr{\la_6 \la_{13}v_\eta^2}{\sqrt{V_{236}^2+V_{236}(\la_3^2 v_\rho^2 -\la_2 v_\eta^2)}}
\right)\, \label{h5aa}
\ee
\be
g_{h_5 A_5 A_5} \approx  \fr{v_\eta^4}{2\sqrt2(v_\eta^2+v_\rho^2)(v_\eta^2+2v_\rho^2)^2} \left(v_\eta(2v_\rho^2 + \la_6 v_\eta^2) \sqrt{\fr{V_{236}+\la_2 v_\eta^2 - \la_3 v_\rho^2}{V_{236}}}
+v_\rho (2 \la_3 v_\eta^2 +\la_6 v_\rho^2) \sqrt{\frac{V_{236}+\la_3 v_\rho^2 - \la_2 v_\eta^2}{V_{236}}}
\right)\,. \label{h5A5A5}
\ee
From  Eq. \eq{haa} to Eq. \eq{h5A5A5}), we can see that couplings $g_{haa}, g_{hA_5A_5}, g_{h_5aa}, g_{h_5A_5A_5}$ depend on $v_\rho, v_\eta$ in EW scale.

\section{Conclusions}

We have analyzed in detail the scalar sector of the 3-3-1 model with ALP. In the model under consideration,  there are two kinds of scalar fields: the bilepton scalars carrying lepton number two and ordinary ones without lepton number. We show that there is no mixing among these two kinds of scalar fields. Moreover, relations among VEVs are related to the self-interactions of scalar fields. 
The physical fields of ALP $a$ and pseudoscalar $A_5$  are defined exactly to help us show that they just interact with exotic quarks in this model with very tiny strength. As a result, ALP is regarded as a candidate of dark matter. Our numerical analysis of the scalar sector allows to successfully accommodate a pseudoscalar $A_5$ with a mass 
ranging from 100 GeV to 1 TeV.  The results are different from the others which have been published before. The $CP$-even scalar sector of the  model was analyzed as well. Its results allow the existence of a non SM scalar boson with mass in a similar range as the pseudoscalar field $A_5$.
 Numerical analysis has shown the constraints on the couplings $\la_2, \la_3, \la_\phi$ with $\tan \al = \fr{v_\eta}{v_\rho}$ and VEVs of scalar fields $\phi, \chi, \eta, \rho$ to raise the new $CP$ even scalar $h_5$ and $CP$ odd scalar $A_5$ with masses in the TeV or subTeV scale.	
 	Furthermore, we	analyzed the consequences of the model in several flavor changing top quark decays, in rare top quark decays, in the leptonic decays of the SM like Higgs boson as well as in the $K^{0}-\bar{K}^{0}$, $B_{d}^{0}-\bar{B}_{d}^{0}$ and $B_{s}^{0}-\bar{B}_{s}^{0}$ meson oscillations. We have found that the model under consideration is consistent with the experimental constraints arising from these processes.

\section*{Acknowledgments}

V. H. Binh and D. T. Huong  acknowledge the financial support of the Vietnam Academy of Science and Technology under grant CSCL05.01/22-23.
AECH has received funding from ANID-Chile FONDECYT 1210378, ANID PIA/APOYO AFB220004 and Milenio-ANID-ICN2019\_044.  H. N. L. is thankful to Van Lang University.

\appendix
\section{Diagonalization of $CP$-odd mass mixing matrix in basis $(I_\phi, I_{\chi}^3, I_\eta^1, I_\rho )$} \label{$CP$o}
Step by step, the matrix $M_{odd}^2$ in (\ref{Modd5}) can be exactly diagonalized by the Euller method.
\begin{enumerate}
	\item 	
	
	In the basis $(I_\eta^1, I_\rho)$, the squared mass matrix has form:
	\bea
	M_{I_{\eta\rho}}^2 = \left(
	\begin{array}{cc}
		-\fr{A}{4 v_\eta^2} & -\fr{A}{4 v_\eta v_\rho} \\
		-\fr{A}{4 v_\eta v_\rho} & -\fr{A}{4 v_\rho^2} \\
	\end{array}
	\right) \label{MIer}
	\eea
	The matrix in \eq{MIer} has 2 eigenvalues which are 0 and $\fr{-A(v_\eta^2+v_\rho^2)}{4v_\eta^2 v_\rho^2}$. This matrix is diagonalized by the matrix below:
	\bea U_{I_{\eta\rho}}= \left(
	\begin{array}{cc}
		-\fr{v_\rho}{v_\eta \sqrt{\fr{v_\rho^2}{v_\eta^2}+1}} & \fr{1}{\sqrt{\fr{v_\rho^2}{v_\eta^2}+1}} \\
		\fr{v_\eta}{v_\rho \sqrt{\fr{v_\eta^2}{v_\rho^2}+1}} & \fr{1}{\sqrt{\fr{v_\eta^2}{v_\rho^2}+1}} \\
	\end{array}
	\right)
	\eea
	Then we receive the  $4 \times 4$ matrix which is used to diagonalize the matrix $M_{odd}^2$ as following:
	\bea
	U_I^1 = \left(
	\begin{array}{cccc}
		1 & 0 & 0 & 0 \\
		0 & 1 & 0 & 0 \\
		0 & 0 &-\fr{v_\rho}{v_\eta \sqrt{\fr{v_\rho^2}{v_\eta^2}+1}} & \fr{1}{\sqrt{\fr{v_\rho^2}{v_\eta^2}+1}}  \\
		0 & 0 &\fr{v_\eta}{v_\rho \sqrt{\fr{v_\eta^2}{v_\rho^2}+1}} & \fr{1}{\sqrt{\fr{v_\eta^2}{v_\rho^2}+1}}  \\
	\end{array}
	\right)\label{u1}
	\eea
	where the mixing angle $\alpha$ is defined by:
	\be \tan \alpha = \fr{v_\eta}{v_\rho}\,.
	\ee
	
	Under the effect of the matrix $U_1$ in (\ref{u1}), the matrix $M_{odd}^2$ becomes:
	\bea
	M_{I_{\rho}^{diag}}^2 = U_I^1 . M_{odd}^2 . (U_I^1)^T =  \left(
	\begin{array}{cccc}
		-\fr{A}{4 v_\phi^2} & -\fr{A}{4 v_\chi v_\phi} &
		0 & -\fr{A \sqrt{\fr{v_\eta^2}{v_\rho^2}+1}}{4
			v_\eta v_\phi} \\
		-\fr{A}{4 v_\chi v_\phi} & -\fr{A}{4 v_\chi^2} &
		0 & -\fr{A \sqrt{\fr{v_\eta^2}{v_\rho^2}+1}}{4
			v_\eta v_\chi} \\
		0 & 0 & 0 & 0 \\
		-\fr{A \sqrt{\fr{v_\eta^2}{v_\rho^2}+1}}{4 v_\eta
			v_\phi} & -\fr{A \sqrt{\fr{v_\eta^2}{v_\rho^2}+1}}{4 v_\eta v_\chi} & 0 & -\fr{A \left(v_\eta^2+v_\rho^2\right)}{4 v_\eta^2 v_\rho^2} \\
	\end{array}
	\right)\label{MIpdiag}
	\eea
	\item Continuously, we consider the $3 \times 3$ mixing matrix in (\ref{MIpdiag}):
	\bea
	M_{I_{33}}^2= \left(
	\begin{array}{ccc}
		-\fr{A}{4 v_\phi^2} & -\fr{A}{4 v_\chi v_\phi} &
		-\fr{A \sqrt{\fr{v_\eta^2}{v_\rho^2}+1}}{4 v_\eta v_\phi} \\
		-\fr{A}{4 v_\chi v_\phi} & -\fr{A}{4 v_\chi^2} &
		-\fr{A \sqrt{\fr{v_\eta^2}{v_\rho^2}+1}}{4 v_\eta v_\chi} \\
		-\fr{A \sqrt{\fr{v_\eta^2}{v_\rho^2}+1}}{4 v_\eta
			v_\phi} & -\fr{A \sqrt{\fr{v_\eta^2}{v_\rho^2}+1}}{4 v_\eta v_\chi} & -\fr{A \left(v_\eta^2+v_\rho^2\right)}{4 v_\eta^2 v_\rho^2} \\
	\end{array}
	\right) \label{MI33}
	\eea
	The matrix $M_{I_{33}}^2$ in \eq{MI33} has got 3 eigenvalues: $0, 0, \fr{-A}{4}\left(\fr{1}{v_\eta^2}+\fr{1}{v_\rho^2}+\fr{1}{v_\chi^2}+\fr{1}{v_\phi^2}\right)$.
	We use the second eigenstate corresponds to the basis $A_3,I^1_\eta$.
	
	In the basis $A_3,I^1_\chi$, the squared mass matrix has form:
	\bea
	M_{I_{A_3 \chi}}^2 = \left(
	\begin{array}{cc}
		-\fr{A}{4 v_\chi^2} & -\fr{A \sqrt{\fr{v_\eta^2}{v_\rho^2}+1}}{4 v_\eta v_\chi} \\
		-\fr{A \sqrt{\fr{v_\eta^2}{v_\rho^2}+1}}{4 v_\eta
			v_\chi} & -\fr{A \left(v_\eta^2+v_\rho^2\right)}{4 v_\eta^2 v_\rho^2} \\
	\end{array}
	\right) \label{MIA3c}
	\eea
	The matrix $M_{I_{A_3 \chi}}^2$ in (\ref{MIA3c}) has 2 eigenvalues: 0 and $\fr{1}{4} A \left(-\fr{1}{v_\eta^2}-\fr{1}{v_\rho^2}-\fr{1}{v_\chi^2}\right)$. This matrix is diagonalized by the matrix below:
	\bea
	U_{A_3 \chi} = \left(
	\begin{array}{cc}
		-\fr{v_\chi \sqrt{\fr{v_\eta^2}{v_\rho^2}+1}}{v_\eta \sqrt{v_\chi^2 \left(\fr{1}{v_\eta^2}+\fr{1}{v_\rho^2}\right)+1}} & \fr{1}{\sqrt{v_\chi^2 \left(\fr{1}{v_\eta^2}+\fr{1}{v_\rho^2}\right)+1}}
		\\
		\fr{v_\eta}{v_\chi \sqrt{\fr{v_\eta^2}{v_\rho^2}+1} \sqrt{\fr{v_\eta^2 v_\rho^2}{v_\chi^2 \left(v_\eta^2+v_\rho^2\right)}+1}}
		& \fr{1}{\sqrt{\fr{v_\eta^2 v_\rho^2}{v_\chi^2
					\left(v_\eta^2+v_\rho^2\right)}+1}} \\
	\end{array}
	\right)
	\eea
	As a result,  we receive the $4 \times 4$ matrix which is used to diagonalize $M_{I_{\phi}^{diag}}^2$  as follows:
	\bea
	U_I^2 = \left(
	\begin{array}{cccc}
		1 & 0 & 0 & 0 \\
		0 & -\fr{v_\chi \sqrt{\fr{v_\eta^2}{v_\rho^2}+1}}{v_\eta \sqrt{v_\chi^2 \left(\fr{1}{v_\eta^2}+\fr{1}{v_\rho^2}\right)+1}} & 0 & \fr{1}{\sqrt{v_\chi^2 \left(\fr{1}{v_\eta^2}+\fr{1}{v_\rho^2}\right)+1}}
		\\
		0 & 0 & 1 & 0 \\
		0 & \fr{v_\eta}{v_\chi \sqrt{\fr{v_\eta^2}{v_\rho^2}+1} \sqrt{\fr{v_\eta^2 v_\rho^2}{v_\chi^2 \left(v_\eta^2+v_\rho^2\right)}+1}}
		& 0 & \fr{1}{\sqrt{\fr{v_\eta^2 v_\rho^2}{v_\chi^2 \left(v_\eta^2+v_\rho^2\right)}+1}} \\
	\end{array}
	\right)\,. \label{u2}
	\eea
	The mixing angle $\theta_3$ is defined by:
	\be
	\tan \theta_3 = \fr{v_\eta}{ v_\chi\sqrt{\fr{v_\eta^2}{v_\rho^2}+1} }\,.
	\ee
	
	Under the effect of the matrix $U_2$ in (\ref{u2}), the matrix $	M_{I_{\eta}^{diag}}^2$ changes into:
	\bea 	M_{I_{\eta\rho}^{diag}}^2 = U_I^2 . 	M_{I_{\rho}^{diag}}^2 (U_I^2)^T =
	\left(
	\begin{array}{cccc}
		-\fr{A}{4 v_\phi^2} & 0 & 0 & -\fr{A \sqrt{\fr{v_\eta^2}{v_\rho^2}+1} \sqrt{\fr{v_\eta^2 v_\rho^2}{v_\chi^2 \left(v_\eta^2+v_\rho^2\right)}+1}}{4 v_\eta v_\phi} \\
		0 & 0 & 0 & 0 \\
		0 & 0 & 0 & 0 \\
		-\fr{A \sqrt{\fr{v_\eta^2}{v_\rho^2}+1}
			\sqrt{\fr{v_\eta^2 v_\rho^2}{v_\chi^2
					\left(v_\eta^2+v_\rho^2\right)}+1}}{4 v_\eta
			v_\phi} & 0 & 0 & \fr{1}{4} A \left(-\fr{1}{v_\eta^2}-\fr{1}{v_\rho^2}-\fr{1}{v_\chi^2}\right) \\
	\end{array}
	\right)\label{MIerdiag}
	\eea
	Next, we consider  the matrix $2 \times 2$ in \eq{MIerdiag} corresponding to the basis $(A_4, I_\phi)$:
	\bea
	M_{I_{22}}= \left(
	\begin{array}{cc}
		-\fr{A}{4 v_\phi^2} & -\fr{A \sqrt{\fr{v_\eta^2}{v_\rho^2}+1} \sqrt{\fr{v_\eta^2 v_\rho^2}{v_\chi^2 \left(v_\eta^2+v_\rho^2\right)}+1}}{4 v_\eta v_\phi} \\
		-\fr{A \sqrt{\fr{v_\eta^2}{v_\rho^2}+1}
			\sqrt{\fr{v_\eta^2 v_\rho^2}{v_\chi^2
					\left(v_\eta^2+v_\rho^2\right)}+1}}{4 v_\eta
			v_\phi} & \fr{-A}{4}  \left(\fr{1}{v_\eta^2}+\fr{1}{v_\rho^2}+\fr{1}{v_\chi^2}\right) \\
	\end{array}
	\right)\label{MI22}
	\eea
	The matrix in \eq{MI22} is a squared mass matrix in basis $(A_4,I_\phi)$ and has got 2 eigenvalues which are 0 and $\fr{-A}{4}  \left(\fr{1}{v_\eta^2}+\fr{1}{v_\rho^2}+\fr{1}{v_\chi^2}+\fr{1}{v_\phi^2}\right)$. The matrix $ M_{I_{22}}$ is diagonalized by the matrix below:
	\bea
	U_{A_4 \phi} = \left(
	\begin{array}{cc}
		-\fr{v_\phi \sqrt{\fr{v_\eta^2}{v_\rho^2}+1}
			\sqrt{\fr{v_\eta^2 v_\rho^2}{v_\chi^2
					\left(v_\eta^2+v_\rho^2\right)}+1}}{v_\eta
			\sqrt{v_\phi^2 \left(\fr{1}{v_\eta^2}+\fr{1}{v_\rho^2}+\fr{1}{v_\chi^2}\right)+1}} &
		\fr{1}{\sqrt{v_\phi^2 \left(\fr{1}{v_\eta^2}+\fr{1}{v_\rho^2}+\fr{1}{v_\chi^2}\right)+1}} \\
		\fr{v_\eta}{v_\phi \sqrt{\fr{v_\eta^2}{v_\rho^2}+1} \sqrt{\fr{v_\eta^2 v_\rho^2}{v_\chi^2 \left(v_\eta^2+v_\rho^2\right)}+1}
			\sqrt{\fr{v_\eta^2 v_\rho^2 v_\chi^2}{v_\phi^2 \left(v_\eta^2 \left(v_\rho^2+v_\chi^2\right)+v_\rho^2 v_\chi^2\right)}+1}}
		& \fr{1}{\sqrt{\fr{v_\eta^2 v_\rho^2 v_\chi^2}{v_\phi^2 \left(v_\eta^2 \left(v_\rho^2+v_\chi^2\right)+v_\rho^2 v_\chi^2\right)}+1}}
		\\
	\end{array}
	\right)
	\eea
	Hence, we receive the  $4 \times 4$ matrix which is used to diagonalized  $M_{I_{\eta\rho}^{diag}}^2$ in the following form:
	\bea
	U_I^3= \left(
	\begin{array}{cccc}
		-\fr{v_\phi \sqrt{\fr{v_\eta^2}{v_\rho^2}+1}
			\sqrt{\fr{v_\eta^2 v_\rho^2}{v_\chi^2
					\left(v_\eta^2+v_\rho^2\right)}+1}}{v_\eta
			\sqrt{v_\phi^2 \left(\fr{1}{v_\eta^2}+\fr{1}{v_\rho^2}+\fr{1}{v_\chi^2}\right)+1}} & 0 &
		0 & \fr{1}{\sqrt{v_\phi^2 \left(\fr{1}{v_\eta^2}+\fr{1}{v_\rho^2}+\fr{1}{v_\chi^2}\right)+1}} \\
		0 & 1 & 0 & 0 \\
		0 & 0 & 1 & 0 \\
		\fr{v_\eta}{v_\phi \sqrt{\fr{v_\eta^2}{v_\rho^2}+1} \sqrt{\fr{v_\eta^2 v_\rho^2}{v_\chi^2 \left(v_\eta^2+v_\rho^2\right)}+1}
			\sqrt{\fr{v_\eta^2 v_\rho^2 v_\chi^2}{v_\phi^2 \left(v_\eta^2 \left(v_\rho^2+v_\chi^2\right)+v_\rho^2 v_\chi^2\right)}+1}}
		& 0 & 0 & \fr{1}{\sqrt{\fr{v_\eta^2 v_\rho^2
					v_\chi^2}{v_\phi^2 \left(v_\eta^2
					\left(v_\rho^2+v_\chi^2\right)+v_\rho^2
					v_\chi^2\right)}+1}} \\
	\end{array}
	\right)\,.\label{u3}
	\eea
	As the mixing angle $\theta_\phi$ is defined as below:
	\be \tan \theta_\phi = \fr{v_\eta}{v_\phi \sqrt{\fr{v_\eta^2}{v_\rho^2}+1} \sqrt{\fr{v_\eta^2 v_\rho^2}{v_\chi^2 \left(v_\eta^2+v_\rho^2\right)}+1}} =\fr{v_{\chi}}{v_\phi \sqrt{1+v_{\chi}^2 \left( \fr{1}{v_\rho^2} + \fr{1}{v_\eta^2}\right)}}\,.
	\ee
	
	Under the effect of the matrix $U_3$ in (\ref{u3}), the matrix $	M_{I_{\eta\rho}^{diag}}^2$ becomes:
	\bea 	M_{I^{diag}}^2 = U_I^3 .	M_{I_{\eta\rho}^{diag}}^2. (U_I^3)^T =
	\left(
	\begin{array}{cccc}
		0 & 0 & 0 & 0 \\
		0 & 0 & 0 & 0 \\
		0 & 0 & 0 & 0 \\
		0 & 0 & 0 & \fr{A}{4}  \left(-\fr{1}{v_\eta^2}-\fr{1}{v_\rho^2}-\fr{1}{v_\chi^2}-\fr{1}{v_\phi^2}\right) \\
	\end{array}
	\right)\label{MIdiag}
	\eea
	\item Finally, the matrix which is used to diagonalize the matrix $M_{odd}^2$ is:
	\bea U_I = U^3_I . U^2_I . U^1_I \,,
	\eea
	and gets the trigonometric form as below:
	\bea U_{Is} = \left(
	\begin{array}{cccc}
		\cos \theta_\phi & -\sin \theta_3 \sin \theta_\phi & - \sin
		\alpha \cos \theta_3 \sin \theta_\phi & -\cos \alpha
		\cos \theta_3 \sin \theta_\phi \\
		0 & \cos \theta_3 & -\sin \alpha \sin \theta_3 & -\cos
		\alpha \sin \theta_3 \\
		0 & 0 & \cos \alpha & -\sin \alpha \\
		-\sin \theta_\phi & - \sin \theta_3 \cos \theta_\phi & - \sin
		\alpha \cos \theta_3 \cos \theta_\phi & -\cos \alpha
		\cos \theta_3 \cos \theta_\phi \\
	\end{array}
	\right)\,.
	\eea

\end{enumerate}

 The $CP$ odd squared mass matrix $M_{odd}^2$ in (\ref{Modd5}) can be exactly diagonalized by the Euler diagonalization method. 
 Then the physical $CP$ odd scalar fields are related with the $CP$ odd scalars in the interaction basis via the following transformation:
\begin{widetext}
	\bea\left(
	\begin{array}{c}
		a\\
		G_{Z^\prime} \\
		G_Z	\\
		A_5	 \\
	\end{array}
	\right)  = 
	\left(
	\begin{array}{cccc}
		\cos \theta_\phi & -\sin \theta_3 \sin \theta_\phi &- \sin
		\al \cos \theta_3 \sin \theta_\phi & -\cos \al
		\cos \theta_3 \sin \theta_\phi \\
		0 & \cos \theta_3 & -\sin \al \sin \theta_3 & -\cos
		\al \sin \theta_3 \\
		0 & 0 & \cos \al & -\sin \al \\
		\sin \theta_\phi & \sin \theta_3 \cos \theta_\phi & \sin
		\al \cos \theta_3 \cos \theta_\phi & \cos \al
		\cos \theta_3 \cos \theta_\phi \\
	\end{array}
	\right)
	\left(
	\begin{array}{c}
		I_\phi\\
		I_{\chi}^3 \\
		I_\rho	\\
		I_\eta^1		 \\
		
	\end{array}
	\right)\,,
	\label{physstatest}
	\eea
\end{widetext}

Note that the mixing matrix has three angles and one parameter which is entered in expression of the  pseudoscalar $A_5$ mass given  in \eq{mA5}.

\section{Diagonalization of $CP$-even mass mixing matrix in basis $(R^1_\eta, R_\rho, R_\chi^3, R_\phi)$} \label{$CP$e}
The matrix $M_R^2$ in (\ref{MR}) is diagonalized by the Hatree - Fock method. It is split into two matrices: $M_{R_0}^2$ - the main contribution and $M_{Rp}^2$ - the perturbation. Those are satisfied the below equation:
\be M_R^2 = M_{R_0}^2 + M_{Rp}^2\,,
\ee
with
\bea M_{R_0}^2= 2\left(
\begin{array}{cccc}
	0 & 0 & 0 & \fr{A}{4 v_\eta v_\phi}+\fr{1}{2} \lambda _{13} v_\eta v_\phi \\
	0 & 0 & 0 & \fr{A}{4 v_\rho v_\phi}+\fr{1}{2} \lambda _{12} v_\rho v_\phi \\
	0 & 0 & 0 & \fr{A}{4 v_\chi v_\phi}+\fr{1}{2} \lambda _{11} v_\chi v_\phi \\
	\fr{A}{4 v_\eta v_\phi}+\fr{1}{2}
	\lambda _{13} v_\eta v_\phi & \fr{A}{4
		v_\rho v_\phi}+\fr{1}{2} \lambda _{12}
	v_\rho v_\phi & \fr{A}{4 v_\chi
		v_\phi}+\fr{1}{2} \lambda _{11} v_\chi
	v_\phi & \lambda _{10} v_\phi^2-\fr{A}{4
		v_\phi^2} \\
\end{array}
\right) \label{MR0}\,,
\eea
and
\bea M_{Rp}^2 =    {2}\left(
\begin{array}{cccc}
	\lambda _2 v_\eta^2-\fr{A}{4 v_\eta^2} &
	\fr{A}{4 v_\eta v_\rho}+\fr{1}{2}
	\lambda _6 v_\eta v_\rho & \fr{\Binh{A}}{4
		v_\eta v_\chi}+\fr{1}{2} \lambda _4
	v_\eta v_\chi & 0 \\
	\fr{A}{4 v_\eta v_\rho}+\fr{1}{2}
	\lambda _6 v_\eta v_\rho & \lambda _3
	v_\rho^2-\fr{A}{4 v_\rho^2} & \fr{A}{4
		v_\rho v_\chi}+\fr{1}{2} \lambda _5
	v_\rho v_\chi & 0 \\
	\fr{A}{4 v_\eta v_\chi}+\fr{1}{2}
	\lambda _4 v_\eta v_\chi & \fr{A}{4
		v_\rho v_\chi}+\fr{1}{2} \lambda _5
	v_\rho v_\chi & \lambda _1 v_\chi^2-\fr{A}{4 v_\chi^2} & 0 \\
	0 & 0 & 0 & 0 \\
\end{array}
\right)\label{MRp}\,.
\eea
In the limits $v_\rho, v_\eta \ll v_\chi \ll v_\phi $, both of $v_\rho$ and $ v_\eta$ can be considered approximately as zero. This makes the main contribution (\ref{MR0}) change into the below matrix:
\bea
M_{R_{00}}^2 \approx \left(
\begin{array}{cccc}
	0 & 0 & 0 & 0 \\
	0 & 0 & 0 & 0 \\
	0 & 0 & 0 &  \lambda _{11} v_\chi
	v_\phi \\
	0 & 0 &  \lambda _{11} v_\chi v_\phi &    {2}\lambda _{10} v_\phi^2 \\
\end{array}
\right) \label{MR00}
\eea
The matrix (\ref{MR00}) is diagonalized by the matrix:
\bea
U_{44}=\left(
\begin{array}{cccc}
	1 & 0 & 0 & 0 \\
	0 & 1 & 0 & 0 \\
	0 & 0 & -\fr{\sqrt{\lambda _{11}^2 v_\chi^2+\lambda
			_{10}^2 v_\phi^2}+\lambda _{10} v_\phi}{\lambda _{11} v_\chi
		\sqrt{\fr{\left(\sqrt{\lambda _{11}^2 v_\chi^2+\lambda _{10}^2 v_\phi^2}+\lambda _{10}
				v_\phi\right){}^2}{\lambda _{11}^2 v_\chi^2}+1}} & \fr{1}{\sqrt{\fr{\left(\sqrt{\lambda _{11}^2
					v_\chi^2+\lambda _{10}^2 v_\phi^2}+\lambda
				_{10} v_\phi\right){}^2}{\lambda _{11}^2
				v_\chi^2}+1}} \\
	0 & 0 & -\fr{\lambda _{10} v_\phi-\sqrt{\lambda
			_{11}^2 v_\chi^2+\lambda _{10}^2 v_\phi^2}}{\lambda _{11} v_\chi
		\sqrt{\fr{\left(\sqrt{\lambda _{11}^2 v_\chi^2+\lambda _{10}^2 v_\phi^2}-\lambda _{10}
				v_\phi\right){}^2}{\lambda _{11}^2 v_\chi^2}+1}} & \fr{1}{\sqrt{\fr{\left(\sqrt{\lambda _{11}^2
					v_\chi^2+\lambda _{10}^2 v_\phi^2}-\lambda
				_{10} v_\phi\right){}^2}{\lambda _{11}^2
				v_\chi^2}+1}} \\
\end{array}
\right) \label{URphi}\,,
\eea
and the diagonalized matrix of main contribution has form as below:
\bea
M_{R_{00}}^2= U_{44}.M_{R0}^2.U_{44}^T=\left(
\begin{array}{cccc}
	0 & 0 & 0 & 0 \\
	0 & 0 & 0 & 0 \\
	0 & 0 &  v_\phi \left(\lambda _{10}
	v_\phi-\sqrt{\lambda _{11}^2 v_\chi^2+\lambda _{10}^2 v_\phi^2}\right) & 0 \\
	0 & 0 & 0 &  v_\phi \left(\sqrt{\lambda
		_{11}^2 v_\chi^2+\lambda _{10}^2 v_\phi^2}+\lambda _{10} v_\phi\right) \\
\end{array}
\right) \label{MR0diag} \,.
\eea
From (\ref{MR0diag}), the squared mass of inflaton is defined by:
\be m_\phi^2 =  v_\phi \left(\sqrt{\lambda _{11}^2
	v_\chi^2+\lambda _{10}^2 v_\phi^2}+\lambda
_{10} v_\phi\right) \approx    {2}\la_{10} v_\phi^2.
\ee
On the other hand, the matrix $U_{44}$ in (\ref{URphi}) can be presented by another form such as:
\bea
U_{R}^1= \left(
\begin{array}{cccc}
	1 & 0 & 0 & 0 \\
	0 & 1 & 0 & 0 \\
	0 & 0 & -\cos \al_ \phi  & \sin \al_ \phi  \\
	0 & 0 & \sin \al_ \phi  & \cos \al_ \phi  \\
\end{array}
\right) \label{URphilg}\,,
\eea
with
\be \tan 2 \al_\phi = \fr{\lambda _{11} v_{\chi }}{\lambda _{10} v_{\phi }} . \label{alphi}
\ee
The perturbartion $M^2_{Rp}$ is effected by the diagonal matrix $U_{R}^1$ in (\ref{URphilg}) so that it has form:
\bea
M^2_{Rp_{44}}= \left(
\begin{array}{cccc}
	2 \lambda _2 v_\eta^2-\fr{A}{2 v_\eta^2} &
	\fr{A}{2 v_\eta v_\rho}+\lambda _6
	v_\eta v_\rho & -\cos \al_\phi
	\left(\fr{A}{2 v_\eta v_\chi}+\lambda _4
	v_\eta v_\chi\right) & \sin \al_\phi
	\left(\fr{A}{2 v_\eta v_\chi}+\lambda _4
	v_\eta v_\chi\right) \\
	\fr{A}{2 v_\eta v_\rho}+\lambda _6
	v_\eta v_\rho & 2 \lambda _3 v_\rho^2-\fr{A}{2 v_\rho^2} & -\cos \al_\phi
	\left(\fr{A}{2 v_\rho v_\chi}+\lambda _5
	v_\rho v_\chi\right) & \sin \al_\phi
	\left(\fr{A}{2 v_\rho v_\chi}+\lambda _5
	v_\rho v_\chi\right) \\
	-\cos \al_\phi \left(\fr{A}{2 v_\eta
		v_\chi}+\lambda _4 v_\eta v_\chi\right) & -\cos \al_\phi \left(\fr{A}{2
		v_\rho v_\chi}+\lambda _5 v_\rho
	v_\chi\right) & -\fr{\cos ^2\al_\phi
		\left(A-4 \lambda _1 v_\chi^4\right)}{2
		v_\chi^2} & \fr{\sin \al_\phi \cos \al_\phi \left(A-4 \lambda _1 v_\chi^4\right)}{2
		v_\chi^2} \\
	\sin \al_\phi \left(\fr{A}{2 v_\eta
		v_\chi}+\lambda _4 v_\eta v_\chi\right) & \sin \al_\phi \left(\fr{A}{2
		v_\rho v_\chi}+\lambda _5 v_\rho
	v_\chi\right) & \fr{\sin \al_\phi \cos
		\al_\phi \left(A-4 \lambda _1 v_\chi^4\right)}{2 v_\chi^2} & -\fr{\sin ^2\al_\phi \left(A-4 \lambda _1 v_\chi^4\right)}{2
		v_\chi^2} \\
\end{array}
\right)\nn\\
\eea
Because $\al_\phi$ is defined by (\ref{alphi}), then $\sin \al_\phi \rightarrow 0$ when $v_\chi \ll v_\phi$ and $\la_{10}>0$. This helps the matrix $M^2_{Rp_{44}}$ reduce an order and can be rewritten like the form after:
\bea
M^2_{Rp_{44}} = \left(
\begin{array}{cccc}
	2 \lambda _2 v_\eta^2-\fr{A}{2 v_\eta^2} &
	\fr{A}{2 v_\eta v_\rho}+\lambda _6
	v_\eta v_\rho & -\cos \al_\phi
	\left(\fr{A}{2 v_\eta v_\chi}+\lambda _4
	v_\eta v_\chi\right) & 0 \\
	\fr{A}{2 v_\eta v_\rho}+\lambda _6
	v_\eta v_\rho & 2 \lambda _3 v_\rho^2-\fr{A}{2 v_\rho^2} & -\cos \al_\phi
	\left(\fr{A}{2 v_\rho v_\chi}+\lambda _5
	v_\rho v_\chi\right) & 0 \\
	-\cos \al_\phi \left(\fr{A}{2 v_\eta
		v_\chi}+\lambda _4 v_\eta v_\chi\right) & -\cos \al_\phi \left(\fr{A}{2
		v_\rho v_\chi}+\lambda _5 v_\rho
	v_\chi\right) & -\fr{\cos ^2\al_\phi
		\left(A-4 \lambda _1 v_\chi^4\right)}{2
		v_\chi^2} & 0 \\
	0 & 0 & 0 & 0 \\
\end{array}
\right) \label{MRp43}
\eea
From (\ref{MRp43}), one gets a $3 \times 3$ matrix below:
\bea
M^2_{Rp_{33}} = 	M^2_{Rp_{33}^0}+	M^2_{Rp_{33}^p} \,,
\eea
with
\bea
M^2_{Rp_{33}^0}= \left(
\begin{array}{ccc}
	0 & 0 & -\cos \al_\phi \left(\fr{A}{2 v_\eta
		v_\chi}+\lambda _4 v_\eta v_\chi\right) \\
	0 & 0 & -\cos \al_\phi \left(\fr{A}{2 v_\rho
		v_\chi}+\lambda _5 v_\rho v_\chi\right) \\
	-\cos \al_\phi \left(\fr{A}{2 v_\eta
		v_\chi}+\lambda _4 v_\eta v_\chi\right) & -\cos \al_\phi \left(\fr{A}{2
		v_\rho v_\chi}+\lambda _5 v_\rho
	v_\chi\right) & -\fr{\cos ^2\al_\phi
		\left(A-4 \lambda _1 v_\chi^4\right)}{2
		v_\chi^2} \\
\end{array}
\right) \label{M330}
\eea
is considered as the main contribution and
\bea
M^2_{Rp_{33}^p} = \left(
\begin{array}{ccc}
	2 \lambda _2 v_\eta^2-\fr{A}{2 v_\eta^2} &
	\fr{A}{2 v_\eta v_\rho}+\lambda _6
	v_\eta v_\rho & 0 \\
	\fr{A}{2 v_\eta v_\rho}+\lambda _6
	v_\eta v_\rho & 2 \lambda _3 v_\rho^2-\fr{A}{2 v_\rho^2} & 0 \\
	0 & 0 & 0 \\
\end{array}
\right) \label{M33p}
\eea
is a perturbation of $M^2_{Rp_{33}}$.\\
Consider the main contribution $M^2_{Rp_{33}^0}$ in the limit $v_\chi \gg v_\rho, v_\eta$, we get $-\fr{\cos \al_\phi  \left(A+\lambda _4
	v_\eta^2 v_\chi^2\right)}{2 v_\eta v_\chi} \rightarrow 0$ then $M^2_{Rp_{33}^0}$ approximately has form:
\bea
M^2_{Rp_{33}^{00}} \approx \left(
\begin{array}{ccc}
	0 & 0 & 0 \\
	0 & 0 & -\cos \al_\phi \left(\fr{A}{2 v_\rho
		v_\chi}+\lambda _5 v_\rho v_\chi\right) \\
	0 & -\cos \al_\phi \left(\fr{A}{2 v_\rho
		v_\chi}+\lambda _5 v_\rho v_\chi\right) & -\fr{\cos ^2\al_\phi \left(A-4 \lambda
		_1 v_\chi^4\right)}{2 v_\chi^2} \\
\end{array}
\right)\,. \label{M3300}
\eea
The matrix $M^2_{Rp_{33}^{00}}$ in (\ref{M3300}) is diagonalized by the following matrix:
\bea
U_{33}= \left(
\begin{array}{ccc}
	1 & 0 & 0 \\
	0 & -\cos \al_3 & \sin \al_3 \\
	0 & \sin \al_3 & \cos \al_3 \\
\end{array}
\right)\,,
\eea
in which $\al_3$ is defined by:
\be  \tan 2\al_3 = \fr{4 v_\chi
	\left(A+2 \la_5 v_\rho^2
	v_\chi^2\right)}{ \cos \alpha
	\phi  \left(A-4 \la_1	v_\chi^4\right)^2}\,. \label{alchi}
\ee
After being diagonalized, $M^2_{Rp_{33}^{00}}$ has form:
\bea M^2_{Rp_{33}^{diag}}= \left(
\begin{array}{ccc}
	0 & 0 & 0 \\
	0 & m_{H_{\chi_1}}^2
	& 0 \\
	0 & 0 &  m_{H_{\chi_2}}^2
	\\
\end{array}
\right)\,,
\eea
with
\be m_{H_{\chi_{1,2}}}^2 = \fr{-\cos \al_\phi }{4 v_\rho v_\chi^2}\left( A v_\rho \cos \al_\phi -4 \lambda _1 v_\rho v_\chi^4 \cos
\al_\phi  \pm \sqrt{4 v_\chi^2 \left(A+2\lambda _5 v_\rho^2 v_\chi^2\right){}^2+\left(A v_\rho \cos \al_\phi -4 \lambda _1 v_\rho v_\chi^4 \cos
	\al_\phi \right){}^2}\right)\,.
\ee
Because of the condition $m_{H_{\chi}}^2 >0$, $v_\phi \gg v_\chi$ and $\la_\phi$ is very tiny then one gets:
\be m_{H_{\chi}}^2=  \left(\lambda _1 v_\chi^2+ v_\chi \sqrt{\lambda _5^2 v_\rho^2+\lambda _1^2 v_\chi^2}\right)    {\approx 2 \la_1 v_\chi^2 + \fr{\la_5^2}{2 \la_1} v_\rho^2}\,.
\label{b1}	\ee
With $U_{33}$, we get the $4 \times 4$ diagonal matrix below:
\bea U_R^2= \left(
\begin{array}{cccc}
	1 & 0 & 0 & 0 \\
	0 & -\cos \al_3 & \sin \al_3 & 0
	\\
	0 & \sin \al_3 & \cos \al_3 & 0 \\
	0 & 0 & 0 & 1 \\
\end{array}
\right)\,.
\eea
Under the effect of $U_{33}$, the pertubation $	M^2_{Rp_{33}^p} $ changes into the following form:
\bea
M^2_{Rp_{33}^{p^\prime}} = 	\left(
\begin{array}{ccc}
	2 \lambda _2 v_\eta^2-\fr{A}{2 v_\eta^2} & -\cos \al_3 \left(\fr{A}{2 v_\eta v_\rho}+\lambda _6 v_\eta v_\rho\right) & \sin \al_3 \left(\fr{A}{2 v_\eta v_\rho}+\lambda _6 v_\eta v_\rho\right) \\
	-\cos \al_3 \left(\fr{A}{2 v_\eta v_\rho}+\lambda _6 v_\eta v_\rho\right) & -\fr{\cos
		^2\al_3 \left(A-4 \lambda _3 v_\rho^4\right)}{2 v_\rho^2} & \fr{\sin \al_3 \cos
		\al_3 \left(A-4 \lambda _3 v_\rho^4\right)}{2 v_\rho^2} \\
	\sin \al_3 \left(\fr{A}{2 v_\eta v_\rho}+\lambda _6 v_\eta v_\rho\right) & \fr{\sin
		\al_3 \cos \al_3 \left(A-4 \lambda _3 v_\rho^4\right)}{2 v_\rho^2} & -\fr{\sin
		^2\al_3 \left(A-4 \lambda _3 v_\rho^4\right)}{2 v_\rho^2} \\
\end{array}
\right)\,.
\eea
With the limit $v_\chi \gg v_\rho, v_\eta$, we get $\sin \al_3 \rightarrow 0$. So that $M^2_{Rp_{33}^{p^\prime}}$ approximately has form:
\bea M^2_{Rp_{33}^{p^\prime}0} = \left(
\begin{array}{ccc}
	2 \lambda _2 v_\eta^2-\fr{A}{2 v_\eta^2} & -\cos \al_3 \left(\fr{A}{2 v_\eta v_\rho}+\lambda _6 v_\eta v_\rho\right) & 0 \\
	-\cos \al_3 \left(\fr{A}{2 v_\eta v_\rho}+\lambda _6 v_\eta v_\rho\right) & -\fr{\cos
		^2\al_3 \left(A-4 \lambda _3 v_\rho^4\right)}{2 v_\rho^2} & 0 \\
	0 & 0 & 0 \\
\end{array}
\right)\,. \label{MRp32}
\eea
From (\ref{MRp32}), one get the $2 \times 2$ matrix below:
\bea 	M^2_{Rp_{22}}= \left(
\begin{array}{cc}
	2 \lambda _2 v_\eta^2-\fr{A}{2 v_\eta^2} & -\cos \al_3 \left(\fr{A}{2 v_\eta v_\rho}+\lambda _6 v_\eta v_\rho\right) \\
	-\cos \al_3 \left(\fr{A}{2 v_\eta v_\rho}+\lambda _6 v_\eta v_\rho\right) & -\fr{\cos
		^2\al_3 \left(A-4 \lambda _3 v_\rho^4\right)}{2 v_\rho^2} \\
\end{array}
\right)\,. \label{MR22}
\eea
Assuming that $\cos \al_3 \approx 1$, the matrix $	M^2_{Rp_{22}}$ in (\ref{MR22}) can be diagonalized by the $2\times 2$ matrix below:
\bea U_{22} = \left(
\begin{array}{cc}
	-\cos \al_2 & \sin \al_2 \\
	\sin \al_2 & \cos \al_2 \\
\end{array}
\right)\,,
\eea
in which we have

\be \tan 2 \al_2= \fr{4 \cos \al_3 v_\eta v_\rho (A + \la_6 v_\eta^2 v_\rho^2)}{A \cos^2 \al_3 v_\eta^2 -A v_\rho^2 + 4 v_\eta^2 v_\rho^2 (\la_2 v_\eta^2 - \la_3 \cos^2 \al_3 v_\rho^2)}\,. \label{al2}
\ee

After being diagonalized, the matrix $M^2_{Rp_{22}}$ has the form:
\bea M^2_{Rp_{22}^{diag}} = \left(
\begin{array}{cc}
	m_{h_5}^2 & 0 \\
	0 & m_{h}^2
\end{array}
\right)\,,
\eea
with 
\bea
m_{h,h_5}^2 &=&\lambda _2 v_\eta^2-\fr{A}{4 v_\eta^2}-\fr{\cos
	^2\al_3 \left(A-4 \lambda _3 v_\rho^4\right)}{4 v_\rho^2}\nn\\
&& \pm \sqrt{\lambda _6 \cos ^2\al_3 \left(A+\lambda _6 v_\eta^2 v_\rho^2\right)+\lambda _2
	v_\eta^4 \left(A-4 \lambda _3 v_\rho^4\right)+\lambda _3 A v_\rho^4+\fr{\left(A \left(v_\eta^2 \cos 2\al_3+v_\eta^2+2 v_\rho^2\right)-8 \lambda _3 v_\eta^2 v_\rho^4 \cos
		^2\al_3-8 \lambda _2 v_\eta^4 v_\rho^2\right){}^2}{64 v_\eta^4 v_\rho^4}}\nn\\
\eea

With the approximations $\cos \al_3 \approx 1$, ones get:
\bea
m_{h,h_5}^2 &=& \lambda _2 v_\eta^2+\lambda _3 v_\rho^2 -\fr{Av^2}{4 v_\eta^2v_\rho^2}\nn\\
&& \pm\fr{1}{4v_\eta v_\rho}\sqrt{16 v_\eta^2 v_\rho^2 \left(\left(\la_2
	v_\eta^2-\la_3 v_\rho^2\right)^2+\la_6^2 v_\eta^2 v_\rho^2\right)+\lambda \phi ^2 v_\chi^2 v_\phi^2 \left(v_\eta^2+v_\rho^2\right)^2+8
	\lambda \phi  v_\eta v_\rho v_\chi v_\phi
	\left(\la_2 v_\eta^4-v_\eta^2 v_\rho^2
	(\la_2+\la_3-2 \la_6)+\la_3 v_\rho^4\right)}\nn\\ \label{mhh5}
\eea

With $U_{22}$, we get the $4 \times 4$ matrix below:
\bea U_R^3= \left(
\begin{array}{cccc}
	-\cos \al_2 & \sin \al_2 & 0 & 0 \\
	\sin \al_2 & \cos \al_2 & 0 & 0 \\
	0 & 0 & 1 & 0 \\
	0 & 0 & 0 & 1 \\
\end{array}
\right)\,.
\eea
Finally, the matrix which is used to diagonalize $M_R^2$ is:
\bea U_R = U_R^3 . U_R^2 . U_R^1 = \left(
\begin{array}{cccc}
	-\cos \al_2 & -\sin \al_2 \cos \al_3 & -\sin \al_2 \sin \al_3 \cos \al_\phi & \sin \al_2 \sin \al_3 \sin \al_\phi \\
	\sin \al_2 & -\cos \al_2 \cos \al_3 & -\cos \al_2 \sin \al_3 \cos \al_\phi & \cos \al_2 \sin \al_3 \sin \al_\phi \\
	0 & \sin \al_3 & -\cos \al_3 \cos \al_\phi & \cos \al_3 \sin \al_\phi \\
	0 & 0 & \sin \al_\phi & \cos \al_\phi \\
\end{array}
\right)\,.
\eea

Note that comparing to the $4 \times 4$ matrix of $CP$-odd sector containing only four parameters
with three massless solutions,    the matrix in \eq{MR} having 10 parameters are not exactly diagonalized. To solve this problem we have used the Hatree-Fock method where some conditions
such as $v_\phi \gg v_\chi \gg v_\rho, v_\eta$, $\la_\phi \ll 1$  and $\sin \al_3 \approx 0$.
As a consequence, derived matrix contains three angles $\al_2,\al_3$ and $\al_\phi$ and three
parameters associated  with masses of new fields $\Phi, H_\chi$ and $h_5$.

\section{Decay rate of  the SM like Higgs boson into a pair of fermions }\label{hffdecay}

\subsection{SM-like Higgs couplings}

We focus on the coupling of SM-like boson $h$ with two ALP $a$ which is a part of $V$ in (\ref{poten3}):
\be
V \supset \mathcal{V}(h, a, a)\,,
\label{Vhaa} \ee
where
\bea
\fr{2\mathcal{V}(h, a, a)}{haa} =
&-&\fr{2 \lambda _2 v_\eta}{\cos^2 2\al} \cos ^2\alpha  \sin \al_2 \cos ^2\theta_3 \sin ^2\theta_\phi-\fr{\lambda _6
	v_\eta}{\cos^2 2\al} \sin ^2\alpha  \sin \al_2 \cos ^2\theta_3 \sin ^2\theta_\phi-\lambda _4 v_\eta \sin\al_2 \sin ^2\theta_3 \sin ^2\theta_\phi\crn
&-&\lambda _{13} \left(v_\eta \sin \al_2 \cos ^2\theta_\phi
+v_\phi \cos ^2\alpha \sec^22\alpha \cos \al_2 \sin\al_3 \sin\al_\phi \cos ^2\theta_3 \sin ^2\theta_\phi\right)\crn
& + & \fr{2 \lambda _3 v_\rho}{\cos^2 2\al} \sin
^2\alpha  \cos \al_2 \cos\al_3 \cos ^2\theta_3 \sin ^2\theta_\phi\crn
&+&\fr{\lambda _6 v_\rho}{\cos^2 2\al} \cos ^2\alpha
\cos \al_2 \cos\al_3 \cos ^2\theta_3 \sin ^2\theta_\phi+\lambda _5 v_\rho \cos \al_2 \cos
\al_3 \sin ^2\theta_3 \sin ^2\theta_\phi\crn
&+&\lambda _{12} v_\rho \cos \al_2 \cos\al_3 \cos ^2\theta_\phi+\fr{\lambda _4 v_\chi}{\cos^2 2\al} \cos ^2\alpha  \cos \al_2 \sin\al_3 \cos\al_\phi \cos ^2\theta_3 \sin
^2\theta_\phi\crn
&+&\fr{\lambda _5 v_\chi}{\cos^2 2\al} \sin ^2\alpha  \cos \al_2 \sin\al_3 \cos\al_\phi \cos ^2\theta_3 \sin ^2\theta_\phi+2 \lambda _1 v_\chi \cos \al_2 \sin\al_3 \cos\al_\phi \sin ^2\theta_3 \sin ^2\theta_\phi\crn
&+&\lambda _{11} v_\chi \cos \al_2 \sin\al_3 \cos\al_\phi \cos ^2\theta_\phi-\fr{\lambda _{12} v_\phi}{\cos^2 2\al} \sin ^2\alpha
\cos \al_2 \sin\al_3 \sin\al_\phi \cos ^2\theta_3 \sin ^2\theta_\phi\crn
&-&\lambda _{11} v_\phi \cos
\al_2 \sin\al_3 \sin\al_\phi \sin ^2\theta_3 \sin ^2\theta_\phi-2 \lambda _{10} v_\phi \cos \al_2
\sin\al_3 \sin\al_\phi \cos ^2\theta_\phi
\,.
\label{Vhaa1}
\eea

In the limits $v_\phi \gg v_\chi \gg v_\rho, v_\eta$ and $\la_\phi \approx 0$, the mixings angles in (\ref{alchi}), (\ref{al2}) approximately get:
\be
\tan \al_3 \approx \fr{\la_5 v_\rho}{\cos \al_\phi v_\chi}\,, \tan 2\al_2 \approx \fr{\la_6 \cos \al_3 v_\eta v_\rho}{\la_2 v_\eta^2 - \la_3 \cos^2\al_3 v_\rho^2}\,. \label{al23}
\ee
Since $\sin\theta_\phi \approx 0$ and $\sin \al_3 \approx 0 $, hence we neglect the terms associated with them. Then
\bea
\mathcal{V}(h,a, a) &\approx& \fr{haa}{2} \cos ^2\theta_\phi \left(\la_{12} v_\rho \cos\al_2\cos \al_3 -\lambda _{13}v_\eta \sin \al_2\right)\crn
&\approx& \fr{haa}{2\sqrt2}v_\rho v_\eta \left(
\fr{\la_6 \la_{12}}{\sqrt{V_{236}^2+(\la_3 v_\rho^2 - \la_2 v_\eta^2)V_{236}}}-\la_{13}\sqrt{V_{236}+\la_3 v_\rho^2 - \la_2 v_\eta^2}
\right)\,, \label{Vhaa2}
\eea
in which, $V_{236}=\sqrt{\left(\la_2 v_\eta^2-\la_3 v_\rho^2\right)^2+\la_6^2 v_\eta^2 v_\rho^2}$.\\
Similarly about the coupling of SM-like boson $h$ with two pseudoscalar $A_5$, with the limits  $v_\phi \gg v_\chi \gg v_\rho, v_\eta$ and $\la_\phi \approx 0$, ones have:
\bea
\mathcal{V}(h,A_5,A_5) &&\approx \fr{hA_5A_5}{2} \cos^2 \theta_\phi \left[\fr{-2 \la_2 v_\eta}{\cos^2 2\al} \cos^2 \al \cos^2 \theta_3  \sin \al_2 - \la_4 v_\eta \sin \al_2 \sin^2 \theta_3\right.\crn
&&\left.+v_\rho \cos \al_2 \cos \al_3 \left(\fr{2\la_3}{\cos^2 2\al} \cos^2 \theta_3  \sin^2 \al+\la_5 \sin^2 \theta_3\right)+\fr{\la_6 \cos^2 \theta_3}{\cos^2 2\al}  \left(v_\rho \cos^2 \al \cos \al_2 \cos \al_3 - v_\eta \sin^2 \al \sin \al_2 \right)\right]\crn
&&\approx  \fr{hA_5A_5}{2\sqrt2} \left(v_\rho(2\la_3 v_\eta^2 +\la_6 v_\rho^2)\sqrt{\fr{V_{236}-\la_3 v_\rho^2 +\la_2 v_\eta^2}{V_{236}}} -v_\eta(2\la_2 v_\rho^2 +\la_6 v_\eta^2)\sqrt{\fr{V_{236}+\la_3 v_\rho^2 -\la_2 v_\eta^2}{V_{236}}}
\right)\,. \label{VhA5A5}
\eea
The new light boson $h_5$ also has couplings with ALP $a$ and pseudoscalar $A_5$. The potential of $(h_5, a, a)$ coupling is:
\bea
\mathcal{V}(h_5, a, a) &\approx& \fr{h_5 a a}{2} \cos^2 \theta_\phi \left(\la_{12} v_\rho \cos \al_3\sin \al_2 + \la_{13} v_\eta \cos \al_2 \right)\crn
&\approx& \fr{h_5 aa}{2\sqrt2}v_\rho \left(\la_{12}\sqrt{V_{236}+\la_3^2 v_\rho^2 -\la_2 v_\eta^2}+\fr{\la_6 \la_{13}v_\eta^2}{\sqrt{V_{236}^2+V_{236}(\la_3^2 v_\rho^2 -\la_2 v_\eta^2)}}
\right)\,. \label{Vh5aa}
\eea
The coupling $(h_5, A_5, A_5)$ is given by:
\bea
&&\mathcal{V}(h_5,A_5,A_5)\crn
&& \approx \fr{h_5A_5A_5}{2} \cos^2 \theta_\phi\left[\fr{2 \lambda _2 v_\eta}{\cos^2 2\al} \cos ^2\alpha  \cos \al_2 \cos ^2\theta_3+\fr{\lambda _6
	v_\eta}{\cos^2 2\al} \sin ^2\alpha  \cos \al_2 \cos ^2\theta_3\right. \crn
&& \left. +\lambda _4 v_\eta \cos \al_2 \sin
^2\theta_3+ \fr{2 \lambda _3 v_\rho}{\cos^2 2\al} \sin ^2\alpha  \sin \al_2 \cos \al_3 \cos ^2\theta_3+\fr{\lambda _6
	v_\rho}{\cos^2 2\al} \cos ^2\alpha  \sin \al_2 \cos \al_3 \cos ^2\theta_3
\right.\crn
&& +\left.\lambda _5 v_\rho \sin \al_2 \cos \al_3 \sin ^2\theta_3
\right]\crn
&& \approx \fr{h_5A_5A_5}{2\sqrt2} \fr{v_\eta^4}{(v_\eta^2+v_\rho^2)(v_\eta^2+2v_\rho^2)^2} \left(v_\eta(2v_\rho^2 + \la_6 v_\eta^2) \sqrt{\fr{V_{236}+\la_2 v_\eta^2 - \la_3 v_\rho^2}{V_{236}}}
+v_\rho (2 \la_3 v_\eta^2 +\la_6 v_\rho^2) \sqrt{\fr{V_{236}+\la_3 v_\rho^2 - \la_2 v_\eta^2}{V_{236}}}
\right)\crn
\label{Vh5A5A5}
\eea

\subsection{SM-like boson $h$ decays to two fermions}
Let us 
 consider the decay:
\be
h (\vec{p})  \rightarrow f(\vec{k}_1) + \tilde{f}(\vec{k}_2)\,,\hs  f= u, d, c, s, \tau, \mu, e\,.
\label{Decayhff} \ee

Amplitude of the above process is given by

\be
M_{fi}(h \rightarrow f \bar{f}) = g_{(h, f,f)} \bar{u}(\vec{k}_1,s_1)  v(\vec{k}_2,s_2)\,.
\label{Mfihff} \ee

Then, the decay rate of $h \rightarrow \bar{f}f$ process is:
\be
\Ga (h \rightarrow \bar{f}f) =  \int d\Ga = \fr{g_{(h,f,f)}^2}{8\pi}m_h\left(1-\fr{4m_f^2}{m_h^2}\right)^{\fr 3{2}}\,.
\label{decayratehff} \ee

Hence
\bea
\Ga (h \rightarrow \bar{e}e) &=& \fr{\cos^2 \al_2 \cos^2 \alpha_3 \fr{m_e^2}{v_\rho^2}}{8\pi} m_h \left(1-\fr{4m_e^2}{m_h^2}\right)^{\fr 3{2}}
\eea

\end{document}